\documentclass[paper,11pt]{article}
\pdfoutput=1

\makeatletter
\@addtoreset{equation}{section}
\makeatother

\usepackage{cite}
\usepackage{epsfig,amsfonts}
\usepackage{tipa}
\usepackage{mathtools}
\usepackage{graphicx}
\usepackage{caption}
\usepackage{subcaption}
\usepackage{amsmath}
\usepackage{amsthm,amssymb}
\usepackage{dsfont}
\usepackage{upgreek}
\usepackage{cancel}
\usepackage{mathrsfs}
\usepackage{hhline}
\usepackage{upgreek}
\usepackage[ruled,vlined]{algorithm2e}
\usepackage[linktocpage=true]{hyperref}
\usepackage[dvipsnames]{xcolor}
\usepackage{musicography}
\usepackage[normalem]{ulem}
\usepackage{enumitem}

\definecolor{purple_nice}{rgb}{0.4,0.2,0.7}
\definecolor{fuel_blue}{RGB}{42,162,185}
\definecolor{YInMn_blue}{RGB}{46, 80, 144}
\definecolor{ultramarine}{RGB}{63, 0, 255}
\definecolor{KLEIN_blue}{rgb}{0, 0.18, 0.65}
\hypersetup{
    colorlinks=true,
    linkcolor=YInMn_blue,
    citecolor=ultramarine,
    filecolor=fuel_blue,
    urlcolor=KLEIN_blue,
}

\newcommand*{\colorboxed}{}
\def\colorboxed#1#{%
  \colorboxedAux{#1}%
}
\newcommand*{\colorboxedAux}[3]{%
  \begingroup
    \colorlet{cb@saved}{.}%
    \color#1{#2}%
    \boxed{%
      \color{cb@saved}%
      #3%
    }%
  \endgroup
}

\def\half{\textstyle{\frac 1 2}}

\def\column#1#2{\left(\begin{array}[c]{cc}%
#1 &\\
#2 &
\end{array}\!\!\!\!\!\!\right)}
\def\matrix#1#2#3#4{\left(\begin{array}[c]{cc}%
#1 & #2\\
#3 & #4
\end{array}\right)}

\def\XXint#1#2#3{{\setbox0=\hbox{$#1{#2#3}{\int}$}
    \vcenter{\hbox{$#2#3$}}\kern-.5\wd0}}

\def\xb{\bar x}

\def\zb{\bar z}
\def\wb{\bar w}
\def\u{V} 

\def\pz{\partial_z}
\def\pzb{\partial_{\bar z}}

\def\pw{\partial_w}
\def\pwb{\partial_{\bar w}}

\def\be{\begin{equation}}
\def\ee{\end{equation}}
\def\bea{\begin{eqnarray}}
\def\eea{\end{eqnarray}}

\newcommand{\e}{e} 

\newcommand{\TTB}{\mbox{$\mathrm{T}\overline{\mathrm{T}}$}}

\newcommand{\kersg}{\mathcal{K}}
\newcommand{\mr}{R_0} 
\newcommand{\M}{\mathcal{M}} 
\newcommand{\cR}{\mathcal{R}} 
\newcommand{\mcR}{\mathcal{R}_0} 
\newcommand{\gammat}{\Gamma_1^{(t)}}
\newcommand{\gammatb}{\bar{\Gamma}_1^{(t)}}
\newcommand{\ctau}{\uptau} 
\newcommand{\x}{y} 

\renewcommand{\Re}{\operatorname{Re}}

\begin{document}
\begin{titlepage}
\title{{Deforming the ODE/IM correspondence with \mbox{$\mathrm{T}\overline{\mathrm{T}}$}}}
\author{Fabrizio Aramini$^{1\musFlat}$, Nicol\`o Brizio$^{1\musQuarter}$, Stefano Negro$^{2\musNatural}$ and Roberto Tateo$^{1\musSharp}$\\[0.3cm]}
\date{\footnotesize{$^1$ Dipartimento di Fisica, Università di Torino and INFN, Sezione di Torino, Via P. Giuria 1, 10125, Torino, Italy \\[0.1cm]$^2$ Center for Cosmology and Particle Physics, New York University, New York,\\ NY 10003, U.S.A.\\[0.3cm] $^{\musFlat}$\texttt{\href{mailto:}{fabrizio.aramini97@gmail.com},}
$^{\musQuarter}$\texttt{\href{mailto:}{nicolo.brizio@unito.it},}
$^{\musNatural}$\texttt{\href{mailto:stefano.negro@nyu.edu}{stefano.negro@nyu.edu},} $^{\musSharp}$\texttt{\href{mailto:roberto.tateo@unito.it}{roberto.tateo@unito.it},}\\}}
\maketitle

\begin{abstract}
The ODE/IM correspondence is an exact link between classical and quantum integrable models.
The primary purpose of this work is to show that it remains valid after  \mbox{$\mathrm{T}\overline{\mathrm{T}}$} perturbation on both sides of the correspondence. 
In particular, we prove that the deformed   Lax pair of the sinh-Gordon model, obtained from the unperturbed one through a dynamical change of coordinates, leads to the same Burgers-type equation governing the quantum spectral flow induced by \mbox{$\mathrm{T}\overline{\mathrm{T}}$}. Our main conclusions have general validity, as the analysis may be easily adapted to all the known  ODE/IM examples involving integrable quantum field theories.
\end{abstract}
\end{titlepage}
\newpage
\tableofcontents
\newpage
\section{Introduction}
\label{sec:intro}
The first instance of the ODE/IM correspondence dates back to 1998 \cite{Dorey:1998pt} when a surprising connection between works on spectral determinants for specific  Sturm-Liouville problems \cite{sibuya1975global, Voros} and the functional approach to conformal field theories (CFTs) \cite{Bazhanov:1994ft, Bazhanov:1996dr, Suzuki:1999rj,  Dorey:1999pv, Dorey:2006an, Dorey:2007zx, Dorey:2009xa} was first established.
The correspondence holds its roots in the fact that seemingly different quantities in the two contexts fulfil the same set of functional relations \cite{Dorey:1998pt} with identical analytic and asymptotic properties. 

Though initially referring only to the CFT vacuum states, the ODE/IM correspondence was later generalised to encompass excited states \cite{Bazhanov:1998wj, Bazhanov:2003ni} and more recently extended to massive integrable quantum field theories based on the $\mathfrak{sl}_2$ algebra \cite{Gaiotto:2009hg, Lukyanov:2010rn}. The correspondence was also extended to systems based on higher-rank algebras, both critical \cite{Dorey:2000ma,Dorey:2006an,Masoero:2015lga,Masoero:2015rcz} and off-critical \cite{Adamopoulou:2014fca,Dorey:2012bx,Negro:2017xwc}. In this work, we focus on systems based on the $\mathfrak{sl}_2$ algebra.
In the off-critical extension of the original results \cite{Lukyanov:2010rn}, the Sturm-Liouville equation is replaced by the  Lax equations of the modified sinh-Gordon model, written as a pair of second-order differential equations coupled through a field-dependent generalised potential (cf. equation \eqref{potential1}, below). 
In this perspective, we are dealing with two, a priori unrelated systems: a classical integrable equation -- the modified sinh-Gordon one -- and a quantum integrable model -- such as the sine- and sinh-Gordon models. The correspondence can then be viewed as an equality between the conserved quantities of these two systems.\cite{Lukyanov:2010rn, Lukyanov:2013wra, Bazhanov:2013cua}. 

The second main ingredient, relevant to the current purposes, is the recent discovery that specific irrelevant perturbations of quantum field theories can be studied using integrable models tools and hydrodynamics-type flow equations \cite{Smirnov:2016lqw, Cavaglia:2016oda}. The observations made in \cite{Smirnov:2016lqw, Cavaglia:2016oda}, triggered a considerable amount of research activity, with applications ranging from simple systems in quantum mechanics to AdS/CFT and nonlinear electrodynamics \cite{Gross:2019ach, Gross:2019uxi, Chakraborty:2020xwo, He:2021dhr, Ebert:2022xfh, Babaei-Aghbolagh:2020kjg, Babaei-Aghbolagh:2022uij}. 
The perturbation involving Zamolodchikov's $\TTB$ operator \cite{Zamolodchikov:2004ce} is arguably the most interesting representative of an infinite tower of irrelevant deformations \cite{Hernandez-Chifflet:2019sua,Camilo:2021gro,Cordova:2021fnr} (see also \cite{Mussardo:1999aj} for some  early results on irrelevant integrable perturbations).
It is related to the Nambu-Goto string \cite{Caselle:2013dra,Beratto:2019bap,Blair:2020ops}, quantum JT gravity \cite{Dubovsky:2017cnj, Cardy:2018sdv, Dubovsky:2018bmo, Iliesiu:2020zld, Okumura:2020dzb, Ebert:2022ehb}, and it possesses compelling interpretation within the AdS/CFT \cite{McGough:2016lol, Giribet:2017imm, Kraus:2018xrn, Taylor:2018xcy, Hartman:2018tkw, Caputa:2019pam} and supersymmetry \cite{Baggio:2018rpv, Chang:2018dge, Chang:2019kiu, Coleman:2019dvf, Ferko:2019oyv, Ebert:2020tuy} frameworks.  
While at the quantum level, an inviscid  Burgers' equation governs the evolution of the $\TTB$-deformed  spectrum, at the classical level, the perturbation turns out to be equivalent to a dynamical change of the space-time coordinates \cite{Conti:2018jho, Conti:2018tca, Conti:2019dxg, Conti:2022egv}.

The main objective of this work is to unify these two research strands by proving the validity of the ODE/IM after $\TTB$ perturbations of both the classical and the quantum sides of the correspondence. In order to do so, we will prove that the conserved energy and momentum of the theory on the $\TTB$-deformed classical side of the correspondence satisfy the same Burgers' equation as the finite-size spectrum on the $\TTB$-deformed quantum side. This fact ensures that if the ODE/IM is valid for the undeformed systems, it will hold for an arbitrary $\TTB$ deformation. We will proceed in steps, first recalling in section \ref{quantum} how a $\TTB$ deformation alters the finite-size spectrum of a quantum integrable field theory and subsequently reviewing in section \ref{classical} the elements of the ODE/IM correspondence relevant for this work. Section \ref{sec:TTBmap} contains the key arguments and result: a proof that the energy and momentum of a $\TTB$-deformed classical field theory satisfy the inviscid Burgers' equation.

In this work we choose to focus on the specific instance of the ODE/IM correspondence analysed in \cite{Lukyanov:2010rn}, involving the quantum sine-Gordon and the classical modified sinh-Gordon models. Consequently, the arguments of section \ref{sec:TTBmap} are adapted to the specificities of the latter. However, the final result has a much broader validity. In particular, our derivation can be used -- with minimal modifications -- to produce a proof of the Burgers' equation for the energy and momentum of a generic, not necessarily integrable, $\TTB$-deformed classical field theory. To our knowledge, the only previous appearance of the Burgers' equation in a classical theory concerns a very specific class of zero-momentum solutions to the sine-Gordon model \cite{Conti:2018tca}.

Finally, our results should be considered as a first step toward the study of irrelevant deformations using the  ODE/IM correspondence as a powerful quantisation tool. 
\section{The quantum sine-Gordon model at finite volume}\label{quantum}
Integrable models and their properties have been 
extensively studied during past decades, and powerful methods developed to determine the finite-size spectrum of an integrable quantum field theory.
One of these is the non-linear integral equation (NLIE) approach \cite{Klumper1991central, KlumperPearce0,  
KlumperPearce, DDV, DDV2, Bazhanov:1996dr, FioravantiDDV, Feverati:1998dt}, which we briefly review in this section before discussing the effects of the $\TTB$ deformation.

\subsection{The non-linear integral equation}
Consider the sine-Gordon quantum field theory defined on a cylinder of radius $R/2\pi$. The complete information on its spectrum can be extracted from the \emph{counting function} $f_{\nu}(\theta)$, solution to the following NLIE
\cite{KlumperPearce0,KlumperPearce, DDV,FioravantiDDV, Feverati:1998dt}
\begin{align}
\label{eq:sGDDV}\notag
f_\nu(\theta) &=\nu(R,k\,|\,\theta) - \int_{\mathcal{C}_1} d\theta' \, \kersg(\theta - \theta') \, \log\left( 1 + e^{-f_\nu(\theta')} \right) \\
&+ \int_{\mathcal{C}_2} d \theta' \, \kersg(\theta -\theta' ) \, \log\left( 1 + e^{f_\nu(\theta')} \right)\,,
\end{align}
where 
\be
\label{eq:sGdriving}
\nu(R, k\,|\,\theta)=2 \pi \imath\, k -\imath\, mR \sinh(\theta) \;,
\ee
is the so-called driving term. In equations \eqref{eq:sGDDV} and \eqref{eq:sGdriving}, $m$ denotes the sine-Gordon soliton mass and $k\in[-1/2,1/2]$ is the quasi-momentum, also known as the {\it twist}, whose role is to select the vacuum \cite{Zamolodchikov:1994uw,Lukyanov:2010rn}.  In \eqref{eq:sGDDV}, the convolution kernel $\kersg(\theta)$ is
\be\label{kernelsh}
\kersg(\theta) 
= \int_\mathbb{R}\frac{d p}{2 \pi}\cos( p\,\theta) \,\frac{ \sinh\left( \pi p \frac{1-\alpha}{2\alpha} \right) }{ 2 \cosh\left( \pi \frac{p}{2} \right) \, \sinh\left( \pi \frac{p}{2\alpha} \right) }\,,
\ee
where $\alpha=\beta^{-2}-1$ and $\beta^2<1$ is the sine-Gordon coupling. 
For technical reasons, as in \cite{Lukyanov:2010rn}, in the following we  shall limit our attention to the range $\beta^2<1/2$, which corresponds to the region $\alpha>1$.

The information on the specific energy eigenstate state under consideration is encoded in the choice of the integration contours $\mathcal{C}_1$ and $\mathcal{C}_2$. For the ground state $\mathcal{C}_1=\mathbb{R}+\imath 0^+=\mathcal{C}_2^*$, while for the excited states the contours $\mathcal{C}_1$ and $\mathcal{C}_2$ encircle a number of singularities $\{\theta_i\}$ of $\log\left(1 +\e^{f_\nu(\theta_i)}\right)$. See \cite{Bazhanov:1996aq, DT, FioravantiDDV, Feverati:1998dt} for more details.

Energy and momentum can be obtained from the counting function from the following expression
\be
\begin{aligned}
E&=m\left(\int_{\mathcal{C}_1} \frac{dy}{2 \pi \imath} \, \sinh(y) \, \log\left( 1 + e^{-f_\nu(y)} \right)- \int_{\mathcal{C}_2} \frac{dy}{2 \pi \imath}  \, \sinh(y) \, \log\left( 1 + e^{f_\nu(y)} \right)\right)\,,\\
P&=m\left(\int_{\mathcal{C}_1} \frac{dy}{2 \pi \imath}  \, \cosh(y) \, \log\left( 1 + e^{-f_\nu(y)} \right)- \int_{\mathcal{C}_2} \frac{dy}{2 \pi \imath}  \, \cosh(y) \, \log\left( 1 + e^{f_\nu(y)} \right)\right)\,.
\end{aligned}
\label{eq:QsG_En_Mom}
\ee
In particular, the momentum can be computed exactly  via the so-called \emph{dilogarithm trick}. Using this, one easily checks that $P(R) = 2\pi p/R$ with $p\in\mathbb{Z}$ (see for example the Lemma in section 7 of \cite{Destri:1994bv}). 

Finally,  in the following  we shall adopt the following specific parametrization \cite{Lukyanov:2010rn}
\be
\label{eqn:def:r}
mR=2\sqrt{\pi}\mathbf{s}^{\alpha+1}\frac{\Gamma\left(1+\frac{1}{2\alpha}\right)}{\Gamma\left(\frac{3}{2}+\frac{1}{2\alpha}\right)}\,,
\ee
where $\mathbf{s}$ is a dimensionless scaling constant.
\subsection{The $\TTB$ deformation} 
At the level of the non-linear integral equation \eqref{eq:sGDDV},  the $\TTB$ perturbation is introduced by implementing the following modification of the convolution kernel \cite{Cavaglia:2016oda}
\be
\label{eq:modkernel}
\kersg(\theta) \xmapsto{\TTB}  \kersg(\theta) - \tau  \frac{m^2}{2 \pi} \cosh(\theta)\,.
\ee
Inserting \eqref{eq:modkernel} into equation \eqref{eq:sGDDV}, after simple manipulations, one finds that $f_\nu^{(\tau)}(\theta)$, the counting function of the deformed theory, still satisfies \eqref{eq:sGDDV}, up to a redefinition of the driving term
\be  
\label{eq:modsGDDV2}
\nu^{(\tau)}=\nu(\mr,k\,|\,\theta - \theta_0) \;,
\ee
with
\begin{align}
\label{eq:defineRtheta}
&R_0^2=\left(R+\tau E(R,\tau)\right)^2-\tau^2 P^2(R,\tau)\,,\\
&\tanh{\theta_0}=\tau\frac{P(R)}{R+\tau E(R,\tau)}\,.
\end{align}
The redefinition \eqref{eq:modsGDDV2} implies the famous inviscid Burgers' flow equation for the finite-volume quantum spectrum of \TTB-deformed field theories \cite{Cavaglia:2016oda, Smirnov:2016lqw}:
\be\label{burgers}
\frac{\partial}{\partial \tau} E(R,\tau) = E(R,\tau) \frac{\partial}{\partial R} E(R,\tau)+\frac{P(R)^2}{R}\,.
\ee
\section{The modified classical sinh-Gordon model}\label{classical}
In this section, we will follow very closely \cite{Lukyanov:2010rn} in the definition of the quantities of interest. 
Let us consider the modified sinh-Gordon (mShG) model, with equation of motion (EoM)
\be
\label{MShG}
\pz\pzb\eta-e^{2\eta}+p(z,s)p(\zb,s)e^{-2\eta}=0\,,
\ee
where the complex coordinates $\mathbf{z}=(z,\zb)$ are dimensionless. 
The function 
\be\label{p(z)}
p(z,s) = z^{2\alpha}-s^{2\alpha}\,,
\ee
is characterized by the pair of parameters $\alpha$ and $s\in\mathbb{R}_{>0}$. Introducing  polar coordinates,
\be
z=\rho \,e^{\imath\varphi}
\,,\quad\zb=\rho\, e^{-\imath\varphi}\,,
\ee
we can describe the field configurations $\eta(\rho,\varphi)$ relevant to the ODE/IM correspondence as those that satisfy the following requirements
\begin{itemize}
    \item[--] $e^{-\eta(\rho,\varphi)}$ is a single-valued, non-zero complex function on the cone $C_{\pi/\alpha}$ with apex angle $\pi/\alpha$ and $L+\bar{L}$ punctures;
    \item[--] the $\rho\rightarrow\infty$ asymptotic behaviour is $e^{-\eta(\rho,\varphi)}\sim \rho^{-\alpha}$;
    \item[--] the $\rho\rightarrow 0$ asymptotic behaviour is $e^{-\eta(\rho,\varphi)}\sim \rho^{-l}$, with $l\in\left[-1/2,1/2\right]$.
\end{itemize}
Through the ODE/IM correspondence, the conserved quantities of the \emph{classical} mShG model, evaluated on the above field configurations, are identified with the \emph{quantum} ones of the sine-Gordon model, such as its energy and momentum \eqref{eq:QsG_En_Mom}. The parameter $l$ is then related to the quasi-momentum in \eqref{eq:sGdriving} as $l=2|k|-1/2$. The number and positions of the $L+\bar{L}$ punctures determine the specific energy eigenstate on the quantum side of the correspondence and are constrained by a system of algebraic equations, the \emph{monodromy-free conditions} of \cite{Bazhanov:2003ni, Fioravanti:2004cz}, also described in \cite{Bazhanov:2013cua}.
In this section we will describe more in detail the construction of the classical conserved charges of the mShG model. These charges can be understood geometrically as integrals along specific contours. We will leverage this perspective in the following section to apply the $\TTB$ deformation on the classical side of the ODE/IM correspondence.
\subsection{Second order linear differential equations}
As it is well known, the mShG equation \eqref{MShG} can be interpreted as the compatibility condition of the linear problem 
\be\label{linearprob} 
\pz \Psi = L_1 \Psi\, , \quad \pzb\Psi=L_2\Psi\,,
\ee
which involves the Lax pair
\be\label{laxpair}
\begin{aligned}
&L_1=- \frac{1}{2} \pz \eta \, \sigma^3 + e^{\vartheta} \left(\sigma^+ e^{\eta} - \sigma^- p(z,s) e^{-\eta} \right)\,, \\
&L_2=\frac{1}{2} \pzb \eta \, \sigma^3 + e^{-\vartheta} \left(\sigma^- e^{\eta} - \sigma^+ p(\zb,s) e^{-\eta} \right)\,.
\end{aligned}\,
\ee
Here $\sigma^3$ and $\sigma^{\pm}$ are the Pauli matrices and $\vartheta$ is the spectral parameter while $\Psi$ in \eqref{linearprob} is a two-dimensional vector.

One can write the general solution of the linear problem \eqref{linearprob} as
\be
    \Psi=\begin{pmatrix}
                          e^{\frac{\vartheta}{2}}e^{\frac{\eta}{2}}\psi \\
                           e^{-\frac{\vartheta}{2}}e^{-\frac{\eta}{2}}(\pz+\pz\eta)\psi
                  \end{pmatrix} = \begin{pmatrix}
                          e^{\frac{\vartheta}{2}}e^{-\frac{\eta}{2}}(\pzb+\pzb\eta)\bar{\psi} \\
                           e^{-\frac{\vartheta}{2}}e^{\frac{\eta}{2}}\bar{\psi}
                  \end{pmatrix}\,,
\ee
where the auxiliary fields $\psi$ and $\bar{\psi}$ solve the Schrödinger-type equations
\be\label{potential1}
\begin{aligned}
&\left(\pz^2-u(\mathbf{z})-e^{2\vartheta} p(z,s)\right)\psi(\mathbf{z})=0\,,\\
&\left(\pzb^2-\bar{u}(\mathbf{z})-e^{-2\vartheta} p(\zb,s)\right)\bar{\psi}(\mathbf{z})=0\,.
\end{aligned}
\ee
The function $u(\mathbf{z})$ depends on the field $\eta$  appearing in \eqref{MShG} as:
\be
\label{eq:uu1}
u({\bf z})=\left(\partial_z\eta({\bf z})\right)^2-\partial_z^2\eta({\bf z})\,, \quad
\bar{u}({\bf z})=\left(\partial_{\bar{z}}\eta({\bf z})\right)^2-\partial_{\bar{z}}^2\eta({\bf z})\,.
\ee

\subsection{Integrals of motion}
\label{eq:CCurrents}
The modified sinh-Gordon equation \eqref{MShG} is related to the unmodified one by a simple change of variables.
In terms of the new coordinates $\mathbf{w} = (w,\wb)$,
\begin{align}
\label{eqn: w(z)}
     \column{dw}{d\wb}=\column{\sqrt{p(z,s)}dz}{\sqrt{p(\zb,s)}d\zb} ,
\end{align}
the field
\begin{align}
\label{eqn:hat(eta)}
     \hat{\eta}({\bf w})= \eta({\bf z}({\bf w})) -\frac{1}{4} \ln{p(z(w),s)p(\bar{z}(\bar{w}),s)}\,,
\end{align}
satisfies the sinh-Gordon equation
\be
\label{eqn:EoMs}
\pw\pwb\hat{\eta}-e^{2\hat\eta}+e^{-2\hat\eta}=0\,.
\ee
From a  WKB-type analysis of \eqref{potential1} \cite{Faddeev:1987ph, Babelon:2003qtg}, one can find the expressions for the integrals of motion ({\bf IM}s) $\mathcal{J}_{2n-1}$ and $\mathcal{\bar{J}}_{2n-1}$ with $n\in\mathbb{N}_{>0}$. Explicitly:
\begin{align}
\label{eqn:JJbar} 
\begin{aligned}
\mathcal{J}_{2n-1}&=\frac{1}{2(2n-1)\sin\left({\frac{\pi(2n-1)}{2\alpha}}\right)}\int\limits_{\Gamma_1} \left(dw \hat{P}_{2n} + d\wb \hat{R}_{2n-2}\right)\,, \\
\bar {\mathcal{J}}_{2n-1}&=\frac{1}{2(2n-1)\sin\left({\frac{\pi(2n-1)}{2\alpha}}\right)}\int\limits_{\bar{\Gamma}_1}\left( d\wb \hat{\bar {P}}_{2n} + dw \hat{\bar R}_{2n-2}\right)\,.
\end{aligned}
\end{align}
The specification of the integration contours $\Gamma_1$ and $\bar{\Gamma}_1$ is of primary importance and will be discussed in a moment. 
The integrands appearing in \eqref{eqn:JJbar} are closed 1-forms, meaning that they satisfy a continuity equation 
\be
\label{eqn:cont}
\pwb{\hat{P}_{2n}}=\pw{\hat{R}_{2n-2}}
\,,\quad\pw{\hat{\bar{P}}_{2n}}=\pwb{\hat{\bar{R}}_{2n-2}}\,.
\ee
The expressions for $\hat P_{2n}$ and $\hat R_{2n-2}$, in terms of $\hat \eta$, are known for generic $n
$ and can be found in \cite{Lukyanov:2010rn}, however the following analysis will only involve $\hat{P}_2$ and $\hat{R}_0$: 
\begin{align}
\begin{aligned}
&\hat{P}_2= \frac{1}{2} \hat{u}
\,,\quad\hat{R}_0=e^{-2\hat{\eta}} -1\,,\\
&\hat{\bar{P}}_2= \frac{1}{2} \hat{\bar{u}}
\,,\quad\hat{\bar{R}}_0=e^{-2\hat{\eta}}-1 \,.
\end{aligned}
\end{align}
The potentials $\hat{u}$ and $\hat{\bar{u}}$ are defined as in \eqref{eq:uu1} with $\eta$ substituted for $\hat{\eta}$.

As a consequence of the continuity equation \eqref{eqn:cont}, these  densities may be written in terms of the stress-energy tensor components $T_2$, $\bar{T}_2$ and $\Theta_0$ of the sinh-Gordon model in dimensionless coordinates:
\be
\label{eqn:stress-energy}
\begin{split}
&T_2(\mathbf{w})=\frac{1}{2} \hat{u}(\mathbf{w})=\hat{P}_2\,, \quad \bar{T}_2(\mathbf{w})= \frac{1}{2} \hat{\bar{u}}(\mathbf{w})=\hat{\bar{P}}_2\,, \\ &\Theta_0(\mathbf{w})=\hat{R}_0=\hat{\bar{R}}_0=e^{-2\hat{\eta}}-1\,.
\end{split}
\ee
Finally, the total energy $E$ and the total momentum $P$ can be expressed as
\be\label{E:P}
E= \mathcal{E} + \bar{\mathcal{E}} \,,\quad 
P = \mathcal{E} - \bar{\mathcal{E}} \,,
\ee
where
\be\label{epsilon}
\begin{aligned} 
    & \mathcal{E}= \frac{\M}{4} \int\limits_{\Gamma_1}\left(dw \, T_2(\mathbf{w}) + d\wb \, \Theta_0(\mathbf{w})\right)\,,\\
    & \bar{\mathcal{E}}= \frac{\M}{4} \int\limits_{\bar{\Gamma}_1} \left(dw \, \Theta_0(\mathbf{w}) + d\wb \, \bar{T}_2(\mathbf{w})\right) \,.
\end{aligned}
\ee
The dimensionful constant $\M$  is the equivalent of the soliton mass $m$, appearing on the ODE/IM correspondence's quantum side.
\subsection{The integration countours $\Gamma_1$ and $\Gamma_2$}
\label{subsec:R}
While the analysis of the modified sinh-Gordon equation \eqref{MShG} is more easily performed in the coordinates $\mathbf{w}$, we need to keep in mind that the properties of the relevant field configurations are established in the coordinates $\mathbf{z}$. Hence, we need to consider $\Gamma_1$ as the image, through the map $\mathbf{w}(\mathbf{z})$ of a fundamental contour $\Gamma_2$.
Roughly speaking,  this is  a regularization of the contour $\gamma_2$ introduced in 
\cite{Lukyanov:2010rn} by a pair of arches around $z=\infty$ (in the following denoted as $c_2^\pm$) starting just above and just below the positive real axis and  ending on the anti-Stokes lines with directions $\varphi=\pi/(2 \alpha +2)$ and
$\varphi=-\pi/(2 \alpha +2)$. The original contour $\gamma_2$ in \cite{Lukyanov:2010rn} started from $+\infty$ just below the real  $z$ axis, wound around the turning point $z=s$ and went back to $+\infty$ just above the real axis, thus encircling the branch cut of $\sqrt{p(z,s)}$ clockwise. As we will mention momentarily, the regularization provided by $\Gamma_2$ is necessary to ensure the convergence of certain quantities relevant for the ODE/IM correspondence.
Notice that, this simple contour-type regularization works well only in the regime $\alpha> 1$ considered in this paper. 
We shall return to this issue in section \ref{subsec:cyl} below.

\begin{figure}[h!]
\centering
\includegraphics[width=6.3cm,height=8.1cm]{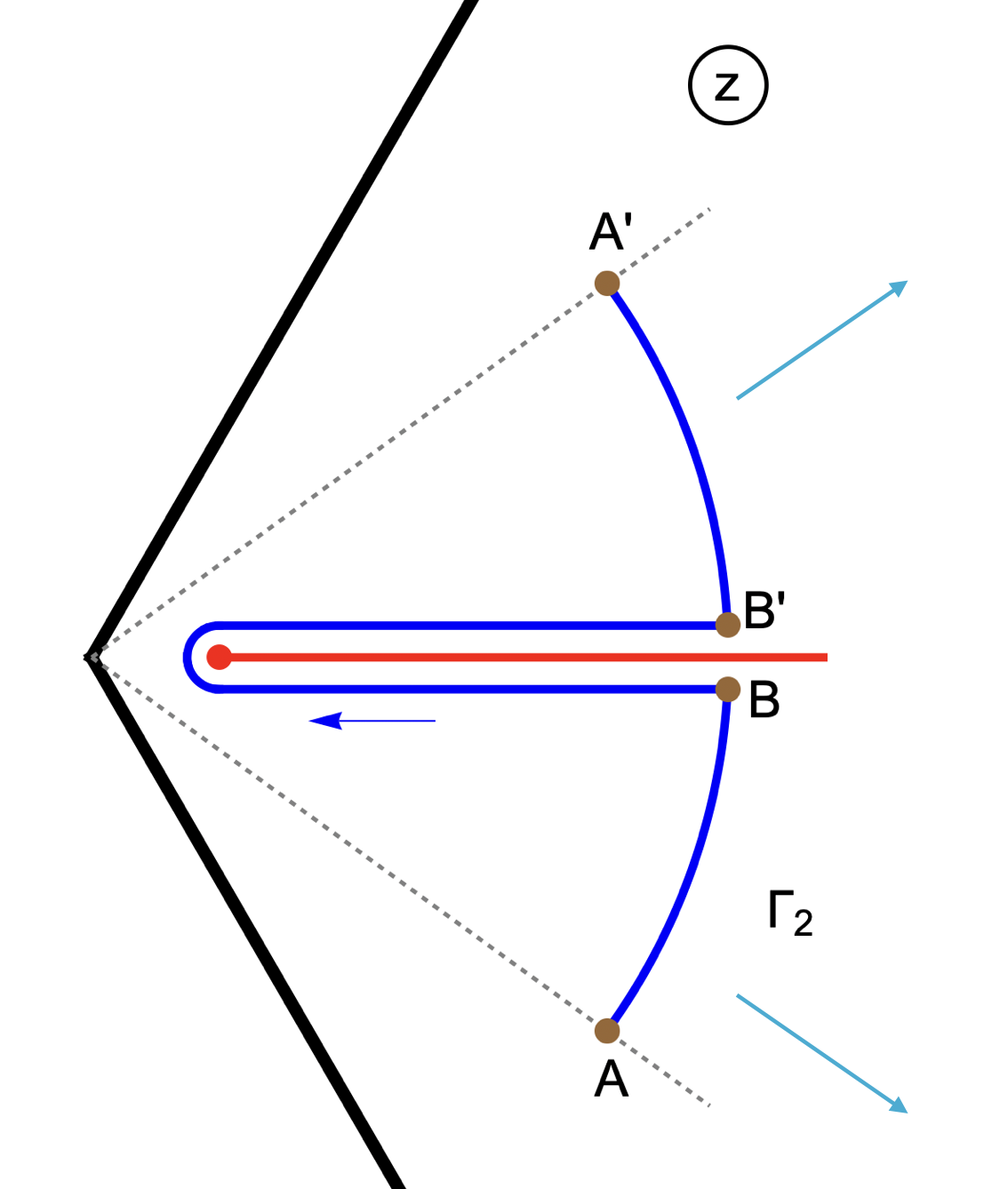}
\caption{\footnotesize The $z$-plane with $\Gamma_2$ in blue. The solid rays $\varphi=\pm \pi/(2\alpha)$ in black and the dashed  rays $\varphi=\pm \pi/(2\alpha+2)$ in gray.}
\label{Gamma2}
\end{figure}

\begin{figure}[h!]
\centering
\includegraphics[width=8.7cm,height=4.5cm]{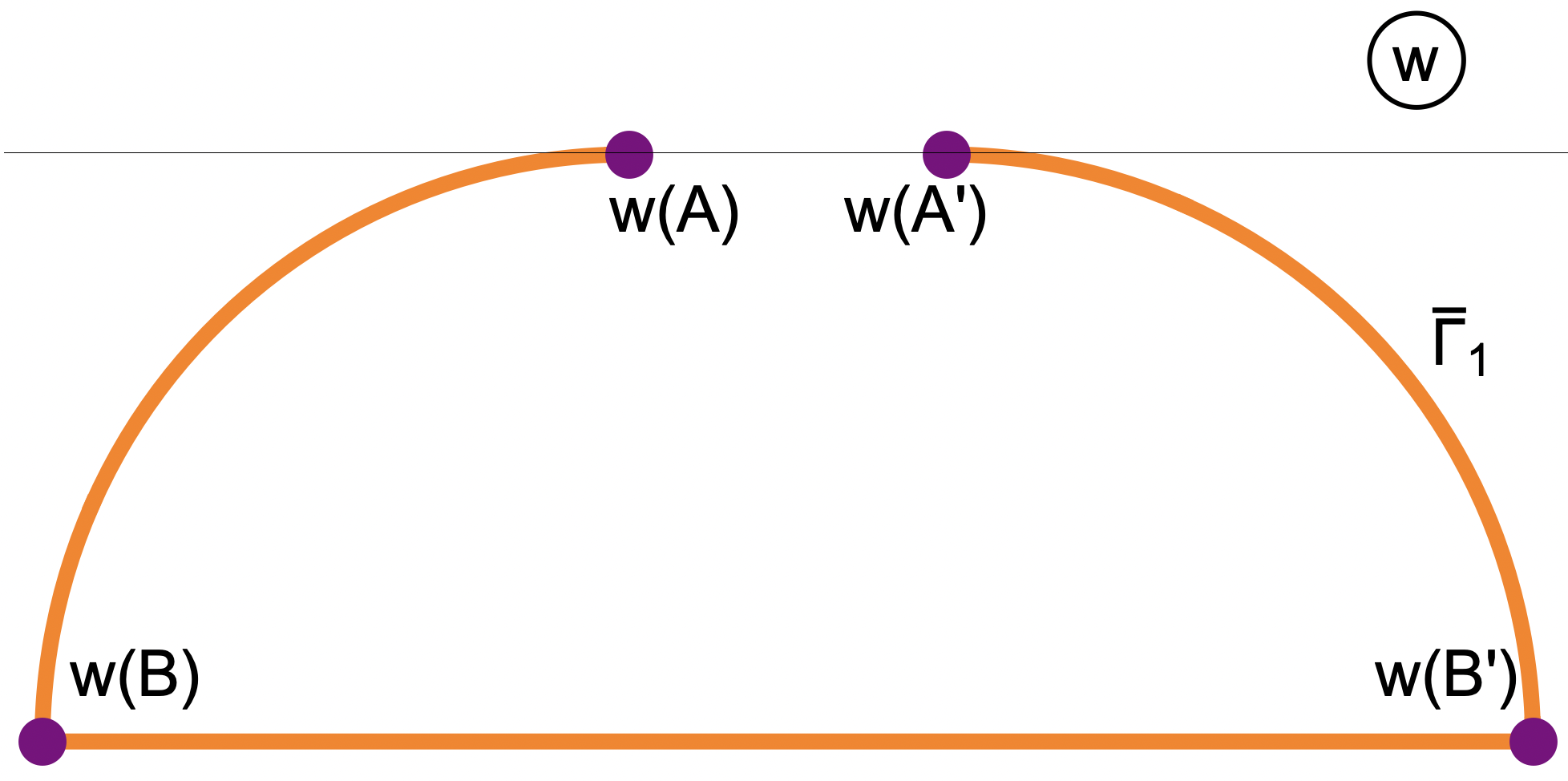}
\caption{\footnotesize The complex $w$-plane with (pictorial) $\bar{\Gamma}_1$ in orange. $w(A')-w(A)=\bar{w}(A)-\bar{w}(A')>0$ is fixed at $\rho\rightarrow+\infty$ (the points $w(B)$ and $w(B')$ are pushed to infinity).
}
\label{Gamma1}
\end{figure}

The contour $\Gamma_2$ is represented in Figure \ref{Gamma2}, where the limit $\rho \rightarrow+\infty$ for the arches' radius is implicitly assumed. Points on the contour $\Gamma_1$ are defined as
\be\label{w:map}
w(P) - w(A) = \int_A^{P \in \Gamma_2} dz\,\sqrt{p(z,s)}\,.
\ee
The qualitative form of the contour $\Gamma_1$, obtained through the map \eqref{w:map}, is represented in Figure \ref{Gamma1}.  Note that, $\bar{\Gamma}_2=\Gamma_2^*$ and $\bar{\Gamma}_1=\Gamma_1^*$. It is important to notice that, in the absence of singularities in the upper part of the $w$-complex plane, $\Gamma_1$ can be straightened to a segment joining the point $w(A)$ to $w(A')$.

Finally,  for integrands with a sufficiently fast vanishing asymptotics for $\rho\rightarrow+\infty$ in the sector  
$ |\arg(z)| \le \pi/(2 \alpha +2)$, the contours $\Gamma_2$ ($\bar \Gamma_2$) and $\gamma_2$ ($\bar \gamma_2$) become totally equivalent.
In particular, given the asymptotic requirements for the field $\eta(z, \bar z)$, in the definition of the \textbf{IM}s $\mathcal{J}_{2n-1}$,  $\bar{\mathcal{J}}_{2n-1}$ one can safely trade the contour $\Gamma_1$ with the image of $\gamma_2$ under the map $\mathbf{w}(\mathbf{z})$. 
\subsection{The circumference of the cylinder}
\label{subsec:cyl}
We shall now introduce the classical equivalent $\M\cR$ of the dimensionless quantity $mR$ appearing in \eqref{eq:sGdriving}. The former is written in terms of simple integrals over $\Gamma_2$ and $\bar{\Gamma}_2$
\be\label{mR:definiton}
\M\cR =\M\cR(s) = \frac{2}{\tan\left(\frac{\pi}{2\alpha}\right)}\int_{\bar{\Gamma}_2}dz\,\sqrt{p(z,s)}=\frac{2}{\tan\left(\frac{\pi}{2\alpha}\right)}\int_{\Gamma_2}d\bar{z}\,\sqrt{p(\bar{z},s)}\,,
\ee
where $\Gamma_2=c_2^-\cup\gamma_2\cup c_2^+$ is represented in Figure \ref{Gamma2}. 
This object can be explicitly computed. First, let us consider the integral appearing in \cite{Lukyanov:2010rn}
\begin{align}
\int_{\bar{\gamma}_2}dz\,\sqrt{p(z,s)}&=2\lim_{\rho\rightarrow+\infty}\int_\rho^sdt\,\sqrt{p(t,s)} \nonumber\\
&=2\lim_{\rho\rightarrow+\infty}\left(\int_\rho^sdt\,\left(\sqrt{p(t,s)}-t^\alpha\right)+\frac{s^{\alpha+1}}{\alpha+1}-\frac{\rho^{\alpha+1}}{\alpha+1}\right) \nonumber \\
&=\tan{\left(\frac{\pi}{2\alpha}\right)}\sqrt{\pi} s^{\alpha+1}\frac{\Gamma\left(1+\frac{1}{2\alpha}\right)}{\Gamma\left(\frac{3}{2}+\frac{1}{2\alpha}\right)}-2\lim_{\rho\rightarrow+\infty}\frac{\rho^{\alpha+1}}{\alpha+1},
\label{eq:int}
\end{align}
where the factor two is due to the discontinuity on the square root type branch cut. This integral is obviously divergent. We choose to regularize it by altering the integration contour. In fact, we calculate the limiting behaviour
\be
\lim_{\rho\rightarrow +\infty} \int_{\bar{c}^+_2(\rho)\,\cup\,\bar{c}^-_2(\rho)} dz\,\sqrt{p(z,s)}=\big(2-2\cos\left(\Omega(\alpha+1)\right)\big)\lim_{\rho\rightarrow +\infty}\frac{\rho^{\alpha+1}}{\alpha+1}.
\ee
Now, if we set $\Omega=\pi/(2\alpha+2)$, the above quantity exactly cancels the divergent term in \eqref{eq:int} and we obtain the following finite expressions
\begin{align}\label{mR:z}
\M\cR(s) =2\sqrt{\pi} s^{\alpha+1}\frac{\Gamma\left(1+\frac{1}{2\alpha}\right)}{\Gamma\left(\frac{3}{2}+\frac{1}{2\alpha}\right)}\,,
\end{align}
establishing a match with the quantity defined in \eqref{eqn:def:r} on the  quantum IM-side of the correspondence: $\M\cR=mR$ or, equivalently,   $s=\mathbf{s}$.

This same result could be obtained by more standard regularization schemes, for example by subtracting the $\rho \rightarrow +\infty$ divergent part appearing on the right-hand side of \eqref{eq:int}, or by  replacing 
\be
\sqrt{p(z,s)}\rightarrow \left(p(z,s)\right)^a\,,
\ee
with $\Re(a) < -1/(2 \alpha)$ and analytically continuing the outcome to $a=1/2$. However, our contour type regularization is more natural, especially in view of the introduction of the $\TTB$ perturbation as a field dependent change of coordinates. 

The same computation in the $\mathbf{w}$ complex plane yields the following simple expression
\begin{align}
\label{mR:reg}
 \M\cR \tan\left(\frac{\pi}{2\alpha}\right) =2\left(\bar{w}(A)-\bar{w}(A')\right)
 = 2\left(w(A')-w(A)\right)\,.
\end{align}
According to the previous definitions \eqref{epsilon} and to the above expression, the total energy and momentum associated to a given field configuration should depend on both $\cR$ and $\M$. However,  an explicit  calculation shows that the momentum is independent of $\M$: $P=P(\cR)$ (see also the discussion in appendix \ref{conformallim}). 

\section{Adding the  $\TTB$}\label{sec:TTBmap}
At the classical level, the simplest way to introduce the $\TTB$ perturbation is through a dynamical change of coordinates that involves the components of the stress-energy tensor \cite{Conti:2019dxg}.
Since we will repeatedly switch from one set of coordinates to the other, we take a moment to fix the notations. The set of coordinates $\bf{w}$ will always be that in which the EoMs \eqref{eqn:EoMs} and, consequently the current densities, look like the unperturbed ones.
In these coordinates, the effect of the $\TTB$ perturbation is implicitly hidden in the integration contour $\Gamma_1^{(t)}$, which  is (equivalent to) the image of  $\Gamma_1$, under a dynamical change of coordinates \cite{Conti:2019dxg}.
On the other hand, in the set of coordinates ${\bf x}$, the effect of the $\TTB$ perturbation will be visible as a modification of both the EoMs and the current densities.
One of the main results of \cite{Conti:2019dxg} is that these two alternative descriptions of a  $\TTB$-perturbed theory are in fact equivalent. This will allow us to prove that the energy of the $\TTB$-deformed, classical mShG model satisfies the Burgers' equation \eqref{burgers}, thus proving the validity of the ODE/IM correspondence for $\TTB$-deformed integrable quantum field theories.
\subsection{The dynamical change of coordinates}
The dynamical change of space-time coordinates  \cite{Conti:2019dxg} is defined as the following differential map:
\be
\column {dw} {d\wb} = J^T \column{dx}{d\xb}\label{eqn:w(x)}\,.
\ee
The Jacobian and its inverse are
\begin{align}
\label{eqn:Jacobian}
\begin{aligned}
& J^T(\mathbf{x})=\frac{1}{\Delta(\mathbf{w}(\mathbf{x}),0)}\matrix{1+2t \Theta_0(\mathbf{w}(\mathbf{x}),0)}{-2t \bar{T}_2(\mathbf{w}(\mathbf{x}),0)}{-2t T_2(\mathbf{w}(\mathbf{x}),0)}{1+2t\Theta_0(\mathbf{w}(\mathbf{x}),0)}\,,\\
& {\left(J^{T}\right)}^{-1}(\mathbf{w},0)= \matrix{1+2t \Theta_0(\mathbf{w},0)}{2t \bar{T}_2(\mathbf{w},0)}{2t T_2(\mathbf{w},0)}{1+2t\Theta_0(\mathbf{w},0)}\,,
\end{aligned}
\end{align}
where $t$ is a dimensionless coupling trivially related to $\tau$ in \eqref{eq:modkernel}, as we will see shortly, while
\be
\label{eqn: Delta}
\Delta(\mathbf{w},0)=(1+2t \Theta_0(\mathbf{w}),0)^2-4t^2T_2(\mathbf{w},0)\bar{T}_2(\mathbf{w},0)\,.
\ee
The explicit form of \eqref{eqn:w(x)} is 
\be \label{w:x}   
\begin{aligned}
&dw=\frac{1}{\Delta(\mathbf{w}(\mathbf{x}),0)}\left(dx+2t \Theta_0(\mathbf{w}(\mathbf{x}),0) dx-2t \bar{T}_2(\mathbf{w}(\mathbf{x}),0)d\bar{x}\right)\,,\\
&d\bar{w}=\frac{1}{\Delta(\mathbf{w}(\mathbf{x}),0)}\left(d\bar{x}+2t \Theta_0(\mathbf{w}(\mathbf{x}),0) d\bar{x}-2t T_2(\mathbf{w}(\mathbf{x}),0)dx\right)\,.\\
\end{aligned}
\ee
Now, after a $\TTB$ deformation, all the quantities introduced in the previous sections will depend on the two independent parameters $\cR$ and $t$.
So for a generic value of $t$, and in agreement with the previous discussions, we can write:
\be
\begin{aligned}
\label{eqn:Perint}
&\mathcal{E}(\cR,t)=\frac{\M}{4} \int\limits_{\gammat}\left(dw \, T_2(\mathbf{w},0) +d\wb \,  \Theta_0(\mathbf{w},0)\right)\,,\\
&\bar{\mathcal{E}}(\cR,t)=\frac{\M}{4} \int\limits_{\gammatb}\left(dw \, \Theta_0(\mathbf{w},0) +d\wb \, \bar{T}_2(\mathbf{w},0) \right)\,,
\end{aligned}
\ee
where, according to \eqref{E:P}:
\be
\label{eq:EP}
E(\cR,t)=\mathcal{E}(\cR,t)+\bar{\mathcal{E}}(\cR,t)\,,\quad P(\cR,t)=\mathcal{E}(\cR,t)-\bar{\mathcal{E}}(\cR,t)\,. 
\ee
Let us stress again that in equations \eqref{eqn:Perint} it is the deformation of the integration contours which drives the evolution in $t$ of the conserved charges.
Notice also how  the map \eqref{eqn:w(x)} implies that the contours $\gammatb$ and $\Gamma_1^{(t)*}$ are not necessarily equivalent, since in general
\be
(T_2(\mathbf{w}),0)^*- \bar{T}_2(\mathbf{w},0)\ne 0\,.
\ee
We shall now shift to the alternative point of view, from which the current densities evolve in $t$ while the contours remain untouched. Using the invariance of $1$-forms under coordinate transformations, we can write
\begin{align}
\begin{aligned}
&T_2(\mathbf{w},0)\,dw  +  \Theta_0(\mathbf{w},0)\,d\wb= T_2(\mathbf{x},t)\,dx + \Theta_0(\mathbf{x},t)\,d\xb\,,\\
&\Theta_0(\mathbf{w},0)\,dw + \bar{T}_2(\mathbf{w},0)\,d\wb = \Theta_0(\mathbf{x},t)\,dx +  \bar{T}_2(\mathbf{x},t)\,d\xb\,,
\end{aligned}
\label{eq:closed_1_forms}
\end{align}
so that
\be
\label{eqn:perturbedstress}
\begin{aligned}
&\mathcal{E}(\cR,t)=\frac{\M}{4} \int\limits_{\Gamma_1} \left(dx \, T_2(\mathbf{x},t) +d\xb \,  \Theta_0(\mathbf{x},t)\right)\,,\\
&\bar{\mathcal{E}}(\cR,t)=\frac{\M}{4} \int\limits_{\bar{\Gamma}_1} \left(dx \,\Theta_0(\mathbf{x},t) +d\xb \, \bar{T}_2(\mathbf{x},t)\right)\,.
\end{aligned}
\ee
Using the explicit form of the map \eqref{w:x} we obtain 
\begin{align}
\label{pert:stressenergy}
\begin{aligned}
&T_2(\mathbf{x},t)=\frac{T_{2}(\mathbf{w}(\mathbf{x}),0)}{\Delta(\mathbf{w}(\mathbf{x}),0)}\,,\quad\bar{T}_2(\mathbf{x},t)= \frac{\bar{T}_{2}(\mathbf{w}(\mathbf{x}),0)}{\Delta(\mathbf{w}(\mathbf{x}),0)}\,,\\
&\Theta_0(\mathbf{x},t)=\frac{\Theta_0(\mathbf{w}(\mathbf{x}),0)+2t(\Theta_0(\mathbf{w}(\mathbf{x}),0)^2-T_2(\mathbf{w}(\mathbf{x}),0)\bar{T}_2(\mathbf{w}(\mathbf{x}),0))}{{\Delta(\mathbf{w}(\mathbf{x}),0})}\,,
\end{aligned}
\end{align}
from which we conclude that
\be\label{w:x_simple}   
\begin{aligned}
&dw=dx - 2t \left( \Theta_0(\mathbf{x},t)\, dx + \bar{T}_2(\mathbf{x},t)\,d\bar{x}\right)\,,\\
&d\bar{w}=d\bar{x}-2t \left(\Theta_0(\mathbf{x},t)\, d\bar{x} + T_2(\mathbf{x},t)\,dx \right)\,.
\end{aligned}
\ee
The inverse map is
\be \label{x:w}   
\begin{aligned}
&dx=dw+2t \left(\Theta_0(\mathbf{w},0)\, dw+
\bar{T}_2(\mathbf{w},0)\,d\bar{w}\right)\,,\\
 &d\bar{x}=d\bar{w}+2t \left(\Theta_0(\mathbf{w},0)\, d\bar{w}+ T_2(\mathbf{w},0)\,dw\right)\,.
 \end{aligned}
\ee

\begin{figure}[t]
  \centering
  \subfloat[]{\includegraphics[scale=0.30]{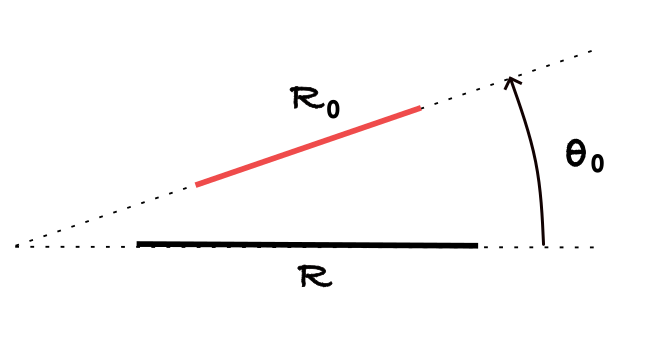}
  \label{fig:Direct}}
  \hspace{1.8cm}
  \subfloat[]{\includegraphics[scale=0.30]{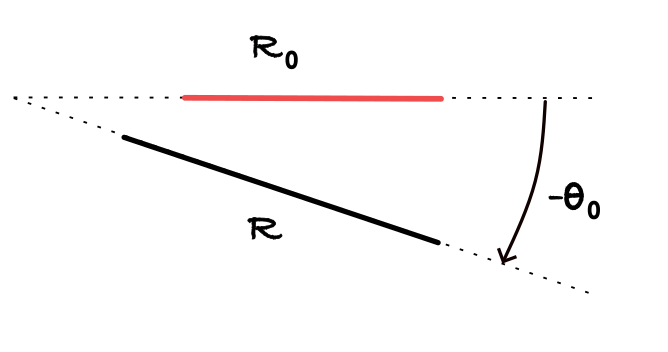}
  \label{fig:Inverse}}
  \vspace{0.1cm}
 \caption{Pictorial representation of the net effect of  the $\TTB$ perturbation, as a rescaling of the `volume' plus a Minkowski rotation. Figure \ref{fig:Direct} corresponds to equation (\ref{eq:LTD}), while Figure \ref{fig:Inverse} to the inverse relation (\ref{eq:LTI}).}
    \label{fig:DI}
\end{figure}

Next, we are going to consider the effect of the $\TTB$ perturbation on the `circumference' $\cR$ \eqref{mR:reg}.  It turns out to be useful to introduce the following parametrization 
\be
\label{eqn:R0}
\M \mcR\, \e^{-\uptheta_0}=\frac{2}{\tan{\left(\frac{\pi}{2\alpha}\right)}} \,\int\limits_{\gammatb} dw \,, \quad \M\mcR \, \e^{+\uptheta_0}=\frac{2}{\tan{\left(\frac{\pi}{2\alpha}\right)}} \,
\int\limits_{\gammat} d\wb \,,
\ee
where $\uptheta_0(t=0) = 0$ and $\mcR(t=0) = \cR$. These expressions will be useful to make contact with the $\TTB$-deformed theory on quantum side of the correspondence \cite{Cavaglia:2016oda}.
 Using the map \eqref{w:x_simple} we write
\begin{align}
\int\limits_{\gammatb} dw &= \int\limits_{\bar{\Gamma}_1}dx -2 t \int\limits_{\bar{\Gamma}_1}\Big(dx\, \Theta_0(\mathbf{x},t)  + d\xb\, \bar{T}_2(\mathbf{x},t) \Big) \nonumber \\
& = \frac{1}{2}\tan\left(\frac{\pi}{2\alpha}\right) \M\cR -  \frac{8}{\M}\, t \bar{\mathcal{E}}(\cR, t)\,,
\end{align}
where relations \eqref{eqn: Delta} and \eqref{pert:stressenergy} were used. Similarly, we obtain
\be
\int\limits_{\gammat} d\wb = \frac{1}{2} \tan{\left(\frac{\pi}{2\alpha}\right)} \M\cR - \frac{8}{\M}\, t \mathcal{E}(\cR, t)\,.
\ee
Setting
\be
\label{eqn: real perturbation parameter}
\ctau= - \frac{8}{ \M^2 \tan{(\frac{\pi}{2\alpha})}} t\,,
\ee
and using relations \eqref{eq:EP} we arrive at 
\begin{align}\label{newhopes}
\begin{cases}
\mcR\,\e^{-\uptheta_0}=\cR+\ctau\left( E(\cR,\tau)-P(\cR,\tau)\right)\\
\mcR\,\e^{+\uptheta_0}=\cR+\ctau\left( E(\cR,\tau)+P(\cR,\tau)\right)
\end{cases}
\,.
\end{align}
The previous analysis and the resulting equations \eqref{newhopes} have the following simple geometric interpretation: the net effect of the $\TTB$ perturbation  is equivalent to a rescaling of $\cR$ plus a Minkowski  rotation
\bea
\cR \rightarrow  \cR_0 \, e^{+\uptheta_0},
\label{eq:LTD}
\eea
with rapidity 
\bea
\uptheta_0 = \frac{1}{2} \log \left(\frac{\cR+\ctau\left(E(\cR,\tau)+P(\cR,\tau)\right)}{\cR+\ctau\left( E(\cR,\tau)-P(\cR,\tau)\right)} \right)\,,
\eea
and
\be
\label{R0} 
\mcR^2=\left(\cR+\ctau E(\cR,\ctau)\right)^2-\ctau^2 P^2(\cR,\ctau)\,,
\ee
as pictorially represented in Figure \ref{fig:Direct}. On the other hand, the dynamical interpretation is encoded in the dependence of $\uptheta_0$ and $\cR_0$ on the energy and momentum of the state.
Therefore, the net effect of the $\TTB$ perturbation corresponds to a  doubling/redefinition of the scaling parameter $s\rightarrow (s^-,s^+)$ of the previous sections. For example, at fixed `mass-scale' $\M$, it corresponds to
\bea
\cR(s) \rightarrow  \left(\cR^+,\cR^- \right)=\left(\cR(s^-),\cR(s^+) \right) = \left(\mcR\,\e^{-\uptheta_0},\mcR\,\e^{+\uptheta_0}\right)\,.
\eea
Notice also that, one might find the interpretation of equations (\ref{newhopes})  more intuitively after switching to the Euclidean version of the theory, obtained through the formal replacements: $P \rightarrow \imath P$ and $\uptheta_0 \rightarrow \imath \uptheta_0$. The interpretation using complex variables is also helpful to guide the intuition in the following final steps of our analysis.
We now need further, independent knowledge to close our system of equations and make it equivalent to the Burgers' equation. Extra constraints can be obtained by demanding the consistency between (\ref{x:w}) and (\ref{w:x_simple}), encoding exact information on the fundamental invertibility property of the Jacobian matrix (\ref{eqn:w(x)}).
Concretely, the  idea is to  derive the analogue of \eqref{newhopes}, for the  inverse transformation 
\bea
\cR_0 \rightarrow \cR \, e^{-\uptheta_0},
\label{eq:LTI}
\eea
as depicted in Figure \ref{fig:Inverse}, by
integrating the r.h.s  of (\ref{x:w}) along appropriate contours $C_1$ and $\bar{C}_1 = C^*_1 $, of a form similar to that of $\Gamma_1$ and $\bar{\Gamma}_1$ (see Figure \ref{Gamma1}).  The endpoints of $\bar{C}_1$  are $w(\tilde{A})$ and $w(\tilde{A}')$, with relative distance
\bea
w(\tilde{A}')-w(\tilde{A}) = \frac{1}{2}\,\M\tan\left(\frac{\pi}{2\alpha}\right)\cR_0\,.
\eea
Under the coordinate transformation $\mathbf{w} \rightarrow \mathbf{x}$, $C_1$ and $\bar C_1$ are mapped to $C_1^{(t)}$  and $\bar{C}_1^{(t)}$, respectively, with
\be \label{eqCC}   
\begin{aligned}
\bar C_1^{(t)}&: \quad \quad x(w(\tilde{A}'))-x(w(\tilde{A})) = \frac{1}{2}\,\M\tan\left(\frac{\pi}{2\alpha}\right)\cR \e^{+\uptheta_0}\,,\\
 C_1^{(t)}&: \quad \quad \bar{x}(\bar{w}(\tilde{A}))-\bar{x}(\bar{w}(\tilde{A}')) = \frac{1}{2}\,\M\tan\left(\frac{\pi}{2\alpha}\right)\cR \e^{-\uptheta_0}\,.
 \end{aligned}
\ee
Proceeding as above, we arrive at the following equations
\begin{align}\label{lasthope}
\begin{cases}
\cR \,\e^{+\uptheta_0}=\mcR-\ctau\left( E(\mcR,0)-P(\mcR,0)\right)\\
\cR\,\e^{-\uptheta_0} =\mcR-\ctau\left( E(\mcR,0)+P(\mcR,0)\right)
\end{cases}\,.
\end{align}
\subsection{The Burgers' equation}
As a final step in our analysis, we shall derive the Burgers' equation involving only classical physical quantities.

From equations \eqref{newhopes} and \eqref{lasthope}, we have
\be\label{sinh:first} 
\sinh{\uptheta_0}= \tau \frac{P(\cR,\uptau)}{\mcR}= \tau \frac{P(\mcR,0)}{\cR}\quad\Longrightarrow \quad\cR P(\cR,\uptau)=\mcR P(\mcR,0)\,,
\ee
implying that  $P(\cR,\uptau)=P(\cR)\propto\cR^{-1}$. Alternatively, the exactly-quantised expression 
\be
P(\cR)= \frac{2 \pi p}{\cR}\,,\quad p\in \mathbb{Z}\,,
\ee
can be obtained through the WKB type analysis reported in appendix \ref{conformallim}.
Putting \eqref{newhopes} and \eqref{lasthope} together, one get
\begin{align}
\begin{cases}
\e^{+\uptheta_0}\,\left( E(\cR,\ctau)-P(\cR,\tau)\right)= E(\mcR,0)-P(\mcR,0)\\
\e^{-\uptheta_0}\,\left( E(\cR,\tau)+P(\cR,\tau)\right)= E(\mcR,0)+P(\mcR,0)\\
\end{cases}
\,,
\end{align}
that is 
\be\label{burgers:impl}    
E^2(\cR,\tau)-P^2(\cR,\tau)=E^2(\mcR,0)-P^2(\mcR,0)\,.
\ee
Moreover, \eqref{newhopes} and \eqref{lasthope} imply the additional constraint \cite{Conti:2019dxg}
\be\label{constraint}   
\frac{\partial}{\partial\ctau}\cR=-E(\cR,\ctau)\,,
\ee
at fixed $\mcR$.
This result together with \eqref{burgers:impl} is the solution, in implicit form, of the Burgers' equation 
\cite{Conti:2018jho}:
\be
\frac{\partial}{\partial \ctau} E(\cR,\ctau) = E(\cR,\ctau) \frac{\partial}{\partial\cR} E(\cR,\ctau)+\frac{P(\cR)^2}{\cR}\,.
\label{eq:class_Burg}
\ee

In conclusion, assuming the validity of the ODE/IM  at $\tau=0$, i.e.
\be
E(R,0)_\textit{quantum}=E(R,0)_\textit{classical}\,,\quad \quad P(R)_\textit{quantum}=P(R)_\textit{classical}\,,
\ee
the fact that the quantum and the classical $\TTB$-evolution equations coincide, implies 
\be
\label{eq:proff}
\colorboxed{purple}{\colorboxed{white}{E(R,\tau)_\textit{quantum}=E(R, \tau)_\textit{classical}}}\,.
\ee
This equality constitutes the key result of the present work.
\section{Conclusions}
The main result of this paper is the extension to $\TTB$ perturbed models of the classical/quantum duality associated with the off-critical variant of the ODE/IM correspondence. Our work links properties of the $\TTB$  perturbation on quantum and classical levels that previously stood on separate grounds.
The analysis highlights a deep connection between the classical coordinate transformation, its invertibility, and the factorization property of the operator $\TTB$, which plays an essential role at the quantum level.
In this work, attention was restricted to the modified sinh-Gordon case of \cite{Lukyanov:2010rn}, however, our result has a much wider validity. First, we notice that the explicit form of the mShG potential \eqref{p(z)} does not appear anywhere in section \ref{sec:TTBmap} -- except for a few constants. Consequently, the validity of the result \eqref{eq:proff} can be immediately extended to ODE/IM correspondence based on more general potentials, \emph{e.g.} the polynomial ones used in the context of AdS/CFT correspondence for the computation of polygonal Wilson loops \cite{Alday:2009dv,Alday:2010vh}. Similarly, it is not difficult to see how the arguments of section \ref{sec:TTBmap} remain valid for $\TTB$ deformations of other known instances of the ODE/IM correspondence \cite{Dorey:2012bx,  Ito:2015nla,  Adamopoulou:2014fca, Lukyanov:2013wra, Bazhanov:2013cua, Ito:2016qzt} and are also easily adaptable to the study of the Lorentz breaking deformations introduced in \cite{Conti:2019dxg}.

We wish to stress that the Burgers' equation \eqref{eq:class_Burg} is actually a general property of $\TTB$-deformed classical field theories. In fact, the argument used in section \ref{sec:TTBmap} teaches us that everything we need to prove \eqref{eq:class_Burg} are the differential maps (\ref{w:x_simple}, \ref{x:w}) and an appropriate choice of Cauchy cycles for both the $\mathbf{w}$ and the $\mathbf{x}$ coordinates,
over which we integrate the closed $1$-forms \eqref{eq:closed_1_forms} in order to obtain the energy and the momentum. Integrating (\ref{w:x_simple}, \ref{x:w}) over these cycles
we arrive at the equivalent of (\ref{newhopes}, \ref{lasthope}), from which the Burgers' equation \eqref{eq:class_Burg} immediately follows.

Another significant point emerging from our analysis is the link between the dynamical change of coordinates and the factorization property of  $\TTB$ \cite{Zamolodchikov:2004ce}. This fact suggests the existence of coordinate transformations also for deformations driven by operators built using higher spin currents. However, preliminary inspections indicate that spacetime maps exist, besides $\TTB$, only for $\mathrm{J}\overline{\mathrm{T}}$ \cite{Guica:2017lia, Chakraborty:2018vja} and the $\mathrm{T}\overline{\mathrm{T}}_s$ Lorentz-breaking models of \cite{Conti:2019dxg},  not for the generic  $\mathrm{T}_s\overline{\mathrm{T}}_{s'}$ deformations. Nevertheless, it may be worth investigating the relation between the latter and maps between the spacetime and more complicated surfaces embedded in higher dimensional spaces. The additional dimensions would correspond to the higher times associated to the commuting Hamiltonian flows generated by the higher conserved charges. A further, natural follow-up  would be the extension of the ODE/IM to the recently discovered $\sqrt{\TTB}$ deformations \cite{Conti:2022egv,Ferko:2022cix, Babaei-Aghbolagh:2022leo, Hou:2022csf, Borsato:2022tmu, Tempo:2022ndz, Garcia:2022wad}.

Finally, it would be nice to use the current framework to understand better the geometric meaning of the Hagedorn singularity appearing in the deformed quantum spectrum and explore at a deeper level the consequences of the $\TTB$ perturbation in the context of minimal surfaces and polygonal Wilson loops \cite{Alday:2009dv,Alday:2010vh} (see also the recent review \cite{Dorey:2019ngq}),  and in the framework of quasi-normal modes of black holes \cite{Gregori:2022xks}.

\section*{Acknowledgements}
This project was partially supported by the INFN project SFT,  grant PHY-2210349, by the Simons Collaboration on Confinement and QCD Strings, and by the FCT Project PTDC/MAT-PUR/30234/2017 ``Irregular connections on algebraic curves and Quantum Field Theory".
S.N. wishes to thank the department of physics of the Universit\`{a} degli Studi di Torino for kind hospitality.

\appendix

\section{The conformal field theory limit}
\label{conformallim}
At the classical level, the ultraviolet limit yielding the conformal field theory behaviour corresponds to the following (left-moving) scaling  limit:
\be\label{CFTlimit}
z=\x\,\e^{-\frac{\vartheta}{1+\alpha}}\,,\quad \vartheta\rightarrow+\infty\,,\quad|z|\sim s\rightarrow0^+\,,
\ee
with
\be
\x=\e^\frac{\vartheta}{1+\alpha} z\,\quad{\bf E}= \e^{2\alpha\vartheta/(1+\alpha)} s^{2\alpha}\,,
\ee
kept finite. In this limit, the first equation in \eqref{potential1} reduced to the Schr\"odinger equation of \cite{Dorey:1998pt}, with  the addition of a finite number of  regular singularities $\left\{\x_a\right\}$ \cite{Bazhanov:1998wj, Bazhanov:2003ni} 
\begin{align}\label{schro:monst}
\left(\frac{d^2}{d\x^2}-\u\left(\x,\{\x_a\}_{a=1}^L\right)-\x^{2\alpha}+{\bf E}\right)\psi_\text{CFT}(\x)=0\,.
\end{align}
Here $\u$ is the so-called `monster potential' \cite{Bazhanov:1998wj},
\be\label{u:monster}
\u\left(\x,\{\x_a\}_{a=1}^L\right)=\frac{l(l+1)}{\x}+2\frac{d^2}{d\x^2}\sum_{a=1}^L\log\left(\x^{2\alpha+2}-\x_a\right)\,,
\ee
and the additional singularities are defined on the cone $\x_a\in C_{\pi/\alpha}\setminus[{\bf E}^{1/(2\alpha)},+\infty)$, away from the apex: $\x_a\ne0$. \\
In the right-moving sector, the opposite scaling limit
\be     
\bar{z}=\x\,\e^{-\frac{\vartheta}{1+\alpha}}\,,  \quad\vartheta\rightarrow-\infty\,,\quad |\bar{z}|\sim s\rightarrow0^+, 
\ee
should be implemented. In this sector, a different number of singularities, $\bar{L}$, will mark the monster potential.

\subsection{WKB analysis and the monster potential}\label{WKB}
A naive, straightforward WKB analysis of \eqref{schro:monst} runs into problems, due to the presence of two competing scales $\textbf{E}$ and $\alpha$. A more refined analysis involves  the following change of variables \cite{Dorey:2004fk}:
\be
\x={\bf E}^{\frac{1}{2\alpha}}\hat{\x}^{\frac{1}{2l+1}}\,,\quad\psi_\text{CFT}(\x)={\bf E}^{-\frac{l}{2\alpha}}\hat{\x}^{-\frac{l}{2l+1}}\hat{\psi}_\text{CFT}(\hat{\x})\,.
\ee
In this new set of coordinates, equation \eqref{schro:monst} reads
\be 
\left(-\epsilon^2\frac{d^2}{d\hat{\x}^2}+q(\hat{\x})+\epsilon^2\xi_L(\hat{\x},\epsilon)\right)\hat{\psi}_\text{CFT}(\hat{\x})=0\,,
\label{eq:ms}
\ee
with
\begin{align}
q(\hat{\x})=\frac{1}{4\lambda^2}\hat{\x}^{\frac{1}{\lambda}}\left(\hat{\x}^{\frac{\alpha}{\lambda}}-1\right)\,,
\end{align}
The last term in the parenthesis, in equation (\ref{eq:ms}), is
\begin{align}
\xi_L(\hat{\x},\epsilon)&=-\frac{1}{\lambda}\frac{d^2}{d\hat{\x}^2}\sum_{a=1}^L\log\left(\hat{\x}^{\frac{\alpha+1}{\lambda}}-\epsilon^2 \x_a\right) \nonumber \\
&+\frac{2\lambda-1}{\lambda^3}(\alpha+1)^2 \hat{\x}^{\frac{\alpha+1}{\lambda}-2}\sum_{a=1}^L\frac{\epsilon^2 \x_a}{\left(\hat{\x}^{\frac{\alpha+1}{\lambda}}-\epsilon^2\x_a\right)^2}\,,
\label{eq:xi}
\end{align}
with
\be 
\lambda=l+\frac{1}{2}\,,\quad\epsilon={\bf E}^{-\frac{\alpha+1}{2\alpha}}\,.
\ee
Now the WKB expansion of $\hat{\psi}_\text{CFT}(\hat{\x})$ is naturally organized by the single parameter $\epsilon$, which combines the two scales $\textbf{E}$ and $\alpha$
\be 
\hat{\psi}_\text{CFT}(\hat{\x})=\text{exp}\left(\frac{1}{\epsilon}\sum_{b\ge0}\epsilon^bS_b(\hat{\x})\right)\,,
\ee
while 
\be
\xi_L(\hat{\x},\epsilon)=\xi_L(\hat{\x},0)+\sum_{c\ge1}\epsilon^c\xi_c^{(L)}(\hat{x})\,.
\ee
The first few terms of the solution are easily computed by recursion,
\begin{align}
&S_0'(\hat{\x})=-\sqrt{q(\hat{\x})}\,,\quad S_1'(\hat{\x})=-\frac{q'(\hat{\x})}{4q(\hat{\x})}\,,\\
&{S_2^{(L)}}'(\hat{\x})=-\frac{1}{48}\left(\frac{q''(\hat{\x})}{q(\hat{\x})^{\frac{3}{2}}}+5\frac{d}{d\hat{\x}}\left(\frac{q''(\hat{\x})}{q(\hat{\x})^{\frac{3}{2}}}\right)\right)-\frac{\xi_L(\hat{\x},0)}{2\sqrt{q(\hat{\x})}}\,.
\end{align}
From $S_2^{(L)}(\hat{\x})$ we can extract  the first {\bf IM} \cite{Dorey:2019ngq} as
\be   
\mathcal{E}=\frac{M}{4}\int_{\gamma_\text{WKB}}d\hat{\x}\,{S_2^{(L)}}'(\hat{\x})\,.
\label{eq:WKBC}
\ee
Here the WKB integration contour $\gamma_\text{WKB}$  starts from $+\infty$ just above the real axis, encircles the turning point $\hat{\x}=1$ and goes back to $+\infty$, just below the real axis. The mass  $M$ sets the energy scale. Similarly, for the right-moving sector
\be   
\bar{\mathcal{E}}=\frac{M}{4}\int_{\gamma_\text{WKB}}d\hat{\x}\,{S_2^{(\bar{L})}}'(\hat{\x})\,.
\ee
Now, we can find the exact expression for the total momentum:
\begin{align}    
P&=\mathcal{E}-\bar{\mathcal{E}}=-\frac{M}{4}\int_{\gamma_\text{WKB}}d\hat{\x}\,\frac{\xi_L(\hat{\x},0)}{2\sqrt{q(\hat{\x})}}+\frac{M}{4}\int_{\gamma_\text{WKB}}d\hat{\x}\,\frac{\xi_{\bar{L}}(\hat{\x},0)}{2\sqrt{q(\hat{\x})}}\nonumber \\
&=\frac{M}{\epsilon}L(\alpha+1)\,\int_{\mathbf{E}^{\frac{1}{2\alpha}}}^{+\infty}\frac{d\x}{\x^2}\frac{1}{\sqrt{\x^{2\alpha}-\mathbf{E}}}-\frac{M}{\epsilon}\bar{L}(\alpha+1)\,\int_{\mathbf{E}^{\frac{1}{2\alpha}}}^{+\infty}\frac{d\x}{\x^2}\frac{1}{\sqrt{\x^{2\alpha}-\mathbf{E}}} \nonumber\\
&=\frac{2\pi(L-\bar{L})}{\cR}\,,
\label{momentum:last}
\end{align}
where, in the last equality in \eqref{momentum:last}, we used the explicit form of $\cR$
\be  
\cR=\frac{2\sqrt{\pi}}{M}\frac{\Gamma\left(1+\frac{1}{2\alpha}\right)}{\Gamma\left(\frac{3}{2}+\frac{1}{2\alpha}\right)}\,.
\ee
Therefore, we arrived at the quantization rule for the total momentum
\be\label{WKB:P}    
P=\frac{2\pi\,p}{\cR}\,,\quad p\in\mathbb{Z}\,,
\ee
Similarly we can compute the total energy, arriving at the following expression
\be\label{WKB:E} 
E=\mathcal{E}+\mathcal{\bar E}=-\frac{\pi}{6\cR}c_\text{eff}\,,
\ee
where $c_\text{eff}$ is the effective central charge:
\be 
c_\text{eff}=1-\frac{6(l+\half)^2}{\alpha+1}-12(L+\bar{L})\,.
\ee 
The results \eqref{WKB:P} and \eqref{WKB:E} can also be derived following other
routes. As an example, we have checked the scaling limit of \cite{Lukyanov:2010rn} (in which $s^{\alpha+1}/\M=M^{-1}$ is kept constant).
From the general assumptions in section \ref{classical}, the WKB analysis reassures us that \eqref{WKB:P} is a rather general result. See also \cite{Conti:2020zft} on recent interesting developments in the study of excited-states in the ODE/IM framework.

Finally, it is worth stressing that the total contribution to the momentum $P$ comes from the region $(z , \bar z) \sim (0,0)$, therefore the result \eqref{WKB:P} is correct also for the full massive version, and after the perturbation with $\TTB$. This is an expected result as, at both the  classical and the quantum levels, the additions of the sin(h)-Gordon potential and $\TTB$ do not modify the original translational properties along the space direction. 

\bibliography{Biblio3}

\providecommand{\href}[2]{#2}\begingroup\raggedright\begin{thebibliography}{10}

\bibitem{Dorey:1998pt}
P.~Dorey and R.~Tateo, {\it {Anharmonic oscillators, the thermodynamic Bethe
  ansatz, and nonlinear integral equations}},  {\em J. Phys.} {\bf A32} (1999)
  L419--L425 [\href{http://arXiv.org/abs/hep-th/9812211}{{\tt
  hep-th/9812211}}].

\bibitem{sibuya1975global}
Y.~Sibuya, {\em Global Theory of a Second Order Linear Ordinary Differential
  Equation with a Polynomial Coefficient}.
\newblock Global Theory of a Second Order Linear Ordinary Differential Equation
  with a Polynomial Coefficient. North-Holland Publishing Company, 1975.

\bibitem{Voros}
A.~Voros, {\it The return of the quartic oscillator. {The} complex {WKB}
  method},  {\em Annales de l'I.H.P. Physique th\'eorique} {\bf 39} (1983),
  no.~3 211--338.

\bibitem{Bazhanov:1994ft}
V.~V. Bazhanov, S.~L. Lukyanov and A.~B. Zamolodchikov, {\it {Integrable
  structure of conformal field theory, quantum KdV theory and thermodynamic
  Bethe ansatz}},  {\em Commun. Math. Phys.} {\bf 177} (1996) 381--398
  [\href{http://arXiv.org/abs/hep-th/9412229}{{\tt hep-th/9412229}}].

\bibitem{Bazhanov:1996dr}
V.~V. Bazhanov, S.~L. Lukyanov and A.~B. Zamolodchikov, {\it {Integrable
  structure of conformal field theory. 2. Q operator and DDV equation}},  {\em
  Commun. Math. Phys.} {\bf 190} (1997) 247--278.

\bibitem{Suzuki:1999rj}
J.~Suzuki, {\it {Anharmonic oscillators, spectral determinant and short exact
  sequence of U(q) (affine sl(2))}},  {\em J. Phys. A} {\bf 32} (1999)
  L183--L188 [\href{http://arXiv.org/abs/hep-th/9902053}{{\tt
  hep-th/9902053}}].

\bibitem{Dorey:1999pv}
P.~Dorey and R.~Tateo, {\it {Differential equations and integrable models: The
  SU(3) case}},  {\em Nucl. Phys. B} {\bf 571} (2000) 583--606
  [\href{http://arXiv.org/abs/hep-th/9910102}{{\tt hep-th/9910102}}]. [Erratum:
  Nucl.Phys.B 603, 582--582 (2001)].

\bibitem{Dorey:2006an}
P.~Dorey, C.~Dunning, D.~Masoero, J.~Suzuki and R.~Tateo, {\it
  {Pseudo-differential equations, and the Bethe ansatz for the classical Lie
  algebras}},  {\em Nucl. Phys.} {\bf B772} (2007) 249--289
  [\href{http://arXiv.org/abs/hep-th/0612298}{{\tt hep-th/0612298}}].

\bibitem{Dorey:2007zx}
P.~Dorey, C.~Dunning and R.~Tateo, {\it {The ODE/IM Correspondence}},  {\em J.
  Phys.} {\bf A40} (2007) R205 [\href{http://arXiv.org/abs/hep-th/0703066}{{\tt
  hep-th/0703066}}].

\bibitem{Dorey:2009xa}
P.~Dorey, C.~Dunning and R.~Tateo, {\it {From PT-symmetric quantum mechanics to
  conformal field theory}},  {\em Pramana} {\bf 73} (2009) 217--239
  [\href{http://arXiv.org/abs/arXiv:0906.1130}{{\tt arXiv:0906.1130}}].

\bibitem{Bazhanov:1998wj}
V.~V. Bazhanov, S.~L. Lukyanov and A.~B. Zamolodchikov, {\it {Spectral
  determinants for Schrodinger equation and Q operators of conformal field
  theory}},  {\em J. Statist. Phys.} {\bf 102} (2001) 567--576
  [\href{http://arXiv.org/abs/hep-th/9812247}{{\tt hep-th/9812247}}].

\bibitem{Bazhanov:2003ni}
V.~V. Bazhanov, S.~L. Lukyanov and A.~B. Zamolodchikov, {\it {Higher level
  eigenvalues of Q operators and Schroedinger equation}},  {\em Adv. Theor.
  Math. Phys.} {\bf 7} (2003), no.~4 711--725
  [\href{http://arXiv.org/abs/hep-th/0307108}{{\tt hep-th/0307108}}].

\bibitem{Gaiotto:2009hg}
D.~Gaiotto, G.~W. Moore and A.~Neitzke, {\it {Wall-crossing, Hitchin systems,
  and the WKB approximation}},  {\em Adv. Math.} {\bf 234} (2013) 239--403
  [\href{http://arXiv.org/abs/arXiv:0907.3987}{{\tt arXiv:0907.3987}}].

\bibitem{Lukyanov:2010rn}
S.~L. Lukyanov and A.~B. Zamolodchikov, {\it {Quantum Sine(h)-Gordon Model and
  Classical Integrable Equations}},  {\em JHEP} {\bf 07} (2010) 008
  [\href{http://arXiv.org/abs/arXiv:1003.5333}{{\tt arXiv:1003.5333}}].

\bibitem{Dorey:2000ma}
P.~Dorey, C.~Dunning and R.~Tateo, {\it {Differential equations for general
  SU(n) Bethe ansatz systems}},  {\em J. Phys. A} {\bf 33} (2000) 8427--8442
  [\href{http://arXiv.org/abs/hep-th/0008039}{{\tt hep-th/0008039}}].

\bibitem{Masoero:2015lga}
D.~Masoero, A.~Raimondo and D.~Valeri, {\it {Bethe Ansatz and the Spectral
  Theory of Affine Lie Algebra-Valued Connections I. The simply-laced Case}},
  {\em Commun. Math. Phys.} {\bf 344} (2016), no.~3 719--750
  [\href{http://arXiv.org/abs/arXiv:1501.07421}{{\tt arXiv:1501.07421}}].

\bibitem{Masoero:2015rcz}
D.~Masoero, A.~Raimondo and D.~Valeri, {\it {Bethe Ansatz and the Spectral
  Theory of Affine Lie algebra\textendash{}Valued Connections II: The Non
  Simply\textendash{}Laced Case}},  {\em Commun. Math. Phys.} {\bf 349} (2017),
  no.~3 1063--1105 [\href{http://arXiv.org/abs/arXiv:1511.00895}{{\tt
  arXiv:1511.00895}}].

\bibitem{Adamopoulou:2014fca}
P.~Adamopoulou and C.~Dunning, {\it {Bethe Ansatz equations for the classical
  $A_n^{(1)}$ affine Toda field theories}},  {\em J. Phys. A} {\bf 47} (2014)
  205205 [\href{http://arXiv.org/abs/arXiv:1401.1187}{{\tt arXiv:1401.1187}}].

\bibitem{Dorey:2012bx}
P.~Dorey, S.~Faldella, S.~Negro and R.~Tateo, {\it {The Bethe Ansatz and the
  Tzitzeica-Bullough-Dodd equation}},  {\em Phil. Trans. Roy. Soc. Lond.} {\bf
  A371} (2013) 20120052 [\href{http://arXiv.org/abs/arXiv:1209.5517}{{\tt
  arXiv:1209.5517}}].

\bibitem{Negro:2017xwc}
S.~Negro, {\it {ODE/IM Correspondence in Toda Field Theories and Fermionic
  Basis in sin(h)-Gordon Model}},  \href{http://arXiv.org/abs/1702.06657}{{\tt
  1702.06657}}.

\bibitem{Lukyanov:2013wra}
S.~L. Lukyanov, {\it {ODE/IM correspondence for the Fateev model}},  {\em JHEP}
  {\bf 12} (2013) 012 [\href{http://arXiv.org/abs/arXiv:1303.2566}{{\tt
  arXiv:1303.2566}}].

\bibitem{Bazhanov:2013cua}
V.~V. Bazhanov and S.~L. Lukyanov, {\it {Integrable structure of Quantum Field
  Theory: Classical flat connections versus quantum stationary states}},  {\em
  JHEP} {\bf 09} (2014) 147 [\href{http://arXiv.org/abs/arXiv:1310.4390}{{\tt
  arXiv:1310.4390}}].

\bibitem{Smirnov:2016lqw}
F.~A. Smirnov and A.~B. Zamolodchikov, {\it {On space of integrable quantum
  field theories}},  {\em Nucl. Phys.} {\bf B915} (2017) 363--383
  [\href{http://arXiv.org/abs/arXiv:1608.05499}{{\tt arXiv:1608.05499}}].

\bibitem{Cavaglia:2016oda}
A.~Cavagli{\`a}, S.~Negro, I.~M. Sz{\'e}c{\'e}snyi and R.~Tateo, {\it {$T
  \bar{T}$-deformed 2D Quantum Field Theories}},  {\em JHEP} {\bf 10} (2016)
  112 [\href{http://arXiv.org/abs/arXiv:1608.05534}{{\tt arXiv:1608.05534}}].

\bibitem{Gross:2019ach}
D.~J. Gross, J.~Kruthoff, A.~Rolph and E.~Shaghoulian, {\it {$T\overline{T}$ in
  AdS$_2$ and Quantum Mechanics}},  {\em Phys. Rev. D} {\bf 101} (2020), no.~2
  026011 [\href{http://arXiv.org/abs/arXiv:1907.04873}{{\tt
  arXiv:1907.04873}}].

\bibitem{Gross:2019uxi}
D.~J. Gross, J.~Kruthoff, A.~Rolph and E.~Shaghoulian, {\it {Hamiltonian
  deformations in quantum mechanics, $T\bar T$, and the SYK model}},  {\em
  Phys. Rev. D} {\bf 102} (2020), no.~4 046019
  [\href{http://arXiv.org/abs/arXiv:1912.06132}{{\tt arXiv:1912.06132}}].

\bibitem{Chakraborty:2020xwo}
S.~Chakraborty and A.~Mishra, {\it {$ T\overline{T} $ and $ J\overline{T} $
  deformations in quantum mechanics}},  {\em JHEP} {\bf 11} (2020) 099
  [\href{http://arXiv.org/abs/arXiv:2008.01333}{{\tt arXiv:2008.01333}}].

\bibitem{He:2021dhr}
S.~He and Z.-Y. Xian, {\it {$T\bar T$ deformation on multiquantum mechanics and
  regenesis}},  {\em Phys. Rev. D} {\bf 106} (2022), no.~4 046002
  [\href{http://arXiv.org/abs/arXiv:2104.03852}{{\tt arXiv:2104.03852}}].

\bibitem{Ebert:2022xfh}
S.~Ebert, C.~Ferko, H.-Y. Sun and Z.~Sun, {\it {$ T\overline{T} $ deformations
  of supersymmetric quantum mechanics}},  {\em JHEP} {\bf 08} (2022) 121
  [\href{http://arXiv.org/abs/arXiv:2204.05897}{{\tt arXiv:2204.05897}}].

\bibitem{Babaei-Aghbolagh:2020kjg}
H.~Babaei-Aghbolagh, K.~B. Velni, D.~M. Yekta and H.~Mohammadzadeh, {\it {$
  T\bar{T} $-like flows in non-linear electrodynamic theories and S-duality}},
  {\em JHEP} {\bf 04} (2021) 187
  [\href{http://arXiv.org/abs/arXiv:2012.13636}{{\tt arXiv:2012.13636}}].

\bibitem{Babaei-Aghbolagh:2022uij}
H.~Babaei-Aghbolagh, K.~B. Velni, D.~M. Yekta and H.~Mohammadzadeh, {\it
  {Emergence of non-linear electrodynamic theories from $T \bar T$-like
  deformations}},  {\em Phys. Lett. B} {\bf 829} (2022) 137079
  [\href{http://arXiv.org/abs/arXiv:2202.11156}{{\tt arXiv:2202.11156}}].

\bibitem{Zamolodchikov:2004ce}
A.~B. Zamolodchikov, {\it {Expectation value of composite field {$T \bar{T}$}
  in two-dimensional quantum field theory}},
  \href{http://arXiv.org/abs/hep-th/0401146}{{\tt hep-th/0401146}}.

\bibitem{Hernandez-Chifflet:2019sua}
G.~Hern\'andez-Chifflet, S.~Negro and A.~Sfondrini, {\it {Flow Equations for
  Generalized $T\overline{T}$ Deformations}},  {\em Phys. Rev. Lett.} {\bf 124}
  (2020), no.~20 200601 [\href{http://arXiv.org/abs/arXiv:1911.12233}{{\tt
  arXiv:1911.12233}}].

\bibitem{Camilo:2021gro}
G.~Camilo, T.~Fleury, M.~Lencs\'es, S.~Negro and A.~Zamolodchikov, {\it {On
  factorizable S-matrices, generalized TTbar, and the Hagedorn transition}},
  {\em JHEP} {\bf 10} (2021) 062
  [\href{http://arXiv.org/abs/arXiv:2106.11999}{{\tt arXiv:2106.11999}}].

\bibitem{Cordova:2021fnr}
L.~C\'ordova, S.~Negro and F.~I. Schaposnik~Massolo, {\it {Thermodynamic Bethe
  Ansatz past turning points: the (elliptic) sinh-Gordon model}},  {\em JHEP}
  {\bf 01} (2022) 035 [\href{http://arXiv.org/abs/arXiv:2110.14666}{{\tt
  arXiv:2110.14666}}].

\bibitem{Mussardo:1999aj}
G.~Mussardo and P.~Simon, {\it {Bosonic type S matrix, vacuum instability and
  CDD ambiguities}},  {\em Nucl. Phys.} {\bf B578} (2000) 527--551
  [\href{http://arXiv.org/abs/hep-th/9903072}{{\tt hep-th/9903072}}].

\bibitem{Caselle:2013dra}
M.~Caselle, D.~Fioravanti, F.~Gliozzi and R.~Tateo, {\it {Quantisation of the
  effective string with TBA}},  {\em JHEP} {\bf 07} (2013) 071
  [\href{http://arXiv.org/abs/arXiv:1305.1278}{{\tt arXiv:1305.1278}}].

\bibitem{Beratto:2019bap}
E.~Beratto, M.~Bill\`o and M.~Caselle, {\it {$T\bar T$ deformation of the
  compactified boson and its interpretation in lattice gauge theory}},  {\em
  Phys. Rev. D} {\bf 102} (2020), no.~1 014504
  [\href{http://arXiv.org/abs/arXiv:1912.08654}{{\tt arXiv:1912.08654}}].

\bibitem{Blair:2020ops}
C.~D.~A. Blair, {\it {Non-relativistic duality and $T \bar T$ deformations}},
  {\em JHEP} {\bf 07} (2020) 069
  [\href{http://arXiv.org/abs/arXiv:2002.12413}{{\tt arXiv:2002.12413}}].

\bibitem{Dubovsky:2017cnj}
S.~Dubovsky, V.~Gorbenko and M.~Mirbabayi, {\it {Asymptotic fragility, near
  AdS$_{2}$ holography and $ T\overline{T} $}},  {\em JHEP} {\bf 09} (2017) 136
  [\href{http://arXiv.org/abs/arXiv:1706.06604}{{\tt arXiv:1706.06604}}].

\bibitem{Cardy:2018sdv}
J.~Cardy, {\it {The $T\overline T$ deformation of quantum field theory as a
  stochastic process}},  \href{http://arXiv.org/abs/arXiv:1801.06895}{{\tt
  arXiv:1801.06895}}.

\bibitem{Dubovsky:2018bmo}
S.~Dubovsky, V.~Gorbenko and G.~Hernandez-Chifflet, {\it {$ T\overline{T} $
  partition function from topological gravity}},  {\em JHEP} {\bf 09} (2018)
  158 [\href{http://arXiv.org/abs/arXiv:1805.07386}{{\tt arXiv:1805.07386}}].

\bibitem{Iliesiu:2020zld}
L.~V. Iliesiu, J.~Kruthoff, G.~J. Turiaci and H.~Verlinde, {\it {JT gravity at
  finite cutoff}},  {\em SciPost Phys.} {\bf 9} (2020) 023
  [\href{http://arXiv.org/abs/arXiv:2004.07242}{{\tt arXiv:2004.07242}}].

\bibitem{Okumura:2020dzb}
S.~Okumura and K.~Yoshida, {\it {$T\bar{T}$-deformation and Liouville
  gravity}},  {\em Nucl. Phys. B} {\bf 957} (2020) 115083
  [\href{http://arXiv.org/abs/arXiv:2003.14148}{{\tt arXiv:2003.14148}}].

\bibitem{Ebert:2022ehb}
S.~Ebert, C.~Ferko, H.-Y. Sun and Z.~Sun, {\it {$T\bar{T}$ in JT Gravity and BF
  Gauge Theory}},  {\em SciPost Phys.} {\bf 13} (2022), no.~4 096
  [\href{http://arXiv.org/abs/arXiv:2205.07817}{{\tt arXiv:2205.07817}}].

\bibitem{McGough:2016lol}
L.~McGough, M.~Mezei and H.~Verlinde, {\it {Moving the CFT into the bulk with $
  T\overline{T} $}},  {\em JHEP} {\bf 04} (2018) 010
  [\href{http://arXiv.org/abs/arXiv:1611.03470}{{\tt arXiv:1611.03470}}].

\bibitem{Giribet:2017imm}
G.~Giribet, {\it {$T\bar{T}$-deformations, AdS/CFT and correlation functions}},
   {\em JHEP} {\bf 02} (2018) 114
  [\href{http://arXiv.org/abs/arXiv:1711.02716}{{\tt arXiv:1711.02716}}].

\bibitem{Kraus:2018xrn}
P.~Kraus, J.~Liu and D.~Marolf, {\it {Cutoff AdS$_{3}$ versus the $
  T\overline{T} $ deformation}},  {\em JHEP} {\bf 07} (2018) 027
  [\href{http://arXiv.org/abs/arXiv:1801.02714}{{\tt arXiv:1801.02714}}].

\bibitem{Taylor:2018xcy}
M.~Taylor, {\it {TT deformations in general dimensions}},
  \href{http://arXiv.org/abs/arXiv:1805.10287}{{\tt arXiv:1805.10287}}.

\bibitem{Hartman:2018tkw}
T.~Hartman, J.~Kruthoff, E.~Shaghoulian and A.~Tajdini, {\it {Holography at
  finite cutoff with a $T^2$ deformation}},  {\em JHEP} {\bf 03} (2019) 004
  [\href{http://arXiv.org/abs/arXiv:1807.11401}{{\tt arXiv:1807.11401}}].

\bibitem{Caputa:2019pam}
P.~Caputa, S.~Datta and V.~Shyam, {\it {Sphere partition functions and cut-off
  AdS}},  {\em JHEP} {\bf 05} (2019) 112
  [\href{http://arXiv.org/abs/arXiv:1902.10893}{{\tt arXiv:1902.10893}}].

\bibitem{Baggio:2018rpv}
M.~Baggio, A.~Sfondrini, G.~Tartaglino-Mazzucchelli and H.~Walsh, {\it {On $
  T\overline{T} $ deformations and supersymmetry}},  {\em JHEP} {\bf 06} (2019)
  063 [\href{http://arXiv.org/abs/arXiv:1811.00533}{{\tt arXiv:1811.00533}}].

\bibitem{Chang:2018dge}
C.-K. Chang, C.~Ferko and S.~Sethi, {\it {Supersymmetry and $ T\overline{T} $
  deformations}},  {\em JHEP} {\bf 04} (2019) 131
  [\href{http://arXiv.org/abs/arXiv:1811.01895}{{\tt arXiv:1811.01895}}].

\bibitem{Chang:2019kiu}
C.-K. Chang, C.~Ferko, S.~Sethi, A.~Sfondrini and G.~Tartaglino-Mazzucchelli,
  {\it {$T\bar{T}$ flows and (2,2) supersymmetry}},  {\em Phys. Rev. D} {\bf
  101} (2020), no.~2 026008 [\href{http://arXiv.org/abs/arXiv:1906.00467}{{\tt
  arXiv:1906.00467}}].

\bibitem{Coleman:2019dvf}
E.~A. Coleman, J.~Aguilera-Damia, D.~Z. Freedman and R.~M. Soni, {\it {$
  T\overline{T} $ -deformed actions and (1,1) supersymmetry}},  {\em JHEP} {\bf
  10} (2019) 080 [\href{http://arXiv.org/abs/arXiv:1906.05439}{{\tt
  arXiv:1906.05439}}].

\bibitem{Ferko:2019oyv}
C.~Ferko, H.~Jiang, S.~Sethi and G.~Tartaglino-Mazzucchelli, {\it {Non-linear
  supersymmetry and $ T\overline{T} $-like flows}},  {\em JHEP} {\bf 02} (2020)
  016 [\href{http://arXiv.org/abs/arXiv:1910.01599}{{\tt arXiv:1910.01599}}].

\bibitem{Ebert:2020tuy}
S.~Ebert, H.-Y. Sun and Z.~Sun, {\it {T$ \overline{T} $ deformation in SCFTs
  and integrable supersymmetric theories}},  {\em JHEP} {\bf 09} (2021) 082
  [\href{http://arXiv.org/abs/arXiv:2011.07618}{{\tt arXiv:2011.07618}}].

\bibitem{Conti:2018jho}
R.~Conti, L.~Iannella, S.~Negro and R.~Tateo, {\it {Generalised Born-Infeld
  models, Lax operators and the $ \mathrm{T}\overline{\mathrm{T}} $
  perturbation}},  {\em JHEP} {\bf 11} (2018) 007
  [\href{http://arXiv.org/abs/arXiv:1806.11515}{{\tt arXiv:1806.11515}}].

\bibitem{Conti:2018tca}
R.~Conti, S.~Negro and R.~Tateo, {\it {The $ \mathrm{T}\overline{\mathrm{T}} $
  perturbation and its geometric interpretation}},  {\em JHEP} {\bf 02} (2019)
  085 [\href{http://arXiv.org/abs/arXiv:1809.09593}{{\tt arXiv:1809.09593}}].

\bibitem{Conti:2019dxg}
R.~Conti, S.~Negro and R.~Tateo, {\it {Conserved currents and
  $\text{T}\bar{\text{T}}_s$ irrelevant deformations of 2D integrable field
  theories}},  {\em JHEP} {\bf 11} (2019) 120
  [\href{http://arXiv.org/abs/arXiv:1904.09141}{{\tt arXiv:1904.09141}}].

\bibitem{Conti:2022egv}
R.~Conti, J.~Romano and R.~Tateo, {\it {Metric approach to a $
  \mathrm{T}\overline{\mathrm{T}} $-like deformation in arbitrary dimensions}},
   {\em JHEP} {\bf 09} (2022) 085
  [\href{http://arXiv.org/abs/arXiv:2206.03415}{{\tt arXiv:2206.03415}}].

\bibitem{Klumper1991central}
A.~Kl{\"u}mper, M.~T. Batchelor and P.~A. Pearce, {\it {Central charges of the
  6- and 19-vertex models with twisted boundary conditions}},  {\em Journal of
  Physics A: Mathematical and General} {\bf 24} (1991), no.~13 3111--3133.

\bibitem{KlumperPearce0}
A.~Kl{\"u}mper and P.~Pearce, {\it {Conformal weights of {RSOS} lattice models
  and their fusion hierarchies}},  {\em Physica A} {\bf 183} (1992) 304.

\bibitem{KlumperPearce}
P.~Pearce and A.~Kl{\"u}mper, {\it {Finite-size corrections and scaling
  dimensions of solvable lattice models: {A}n analytic method}},  {\em Phys.
  Rev. Lett.} {\bf 66} (1991) 974--977.

\bibitem{DDV}
C.~Destri and H.~de~Vega, {\it {New thermodynamic {B}ethe {A}nsatz equations
  without strings}},  {\em Phys. Rev. Lett.} {\bf 69} (Oct, 1992) 2313--2317.

\bibitem{DDV2}
C.~Destri and H.~{De Vega}, {\it {Unified approach to thermodynamic {B}ethe
  {A}nsatz and finite size corrections for lattice models and field theories}},
   {\em Nuclear Physics B} {\bf 438} (1995), no.~3 413--454.

\bibitem{FioravantiDDV}
D.~Fioravanti, A.~Mariottini, E.~Quattrini and F.~Ravanini, {\it {Excited state
  {D}estri-{D}e {V}ega equation for {S}ine-{G}ordon and restricted
  {S}ine-{G}ordon models}},  {\em Phys. Lett. B} {\bf 390} (1997) 243--251
  [\href{http://arXiv.org/abs/hep-th/9608091}{{\tt hep-th/9608091}}].

\bibitem{Feverati:1998dt}
G.~Feverati, F.~Ravanini and G.~Takacs, {\it {Nonlinear integral equation and
  finite volume spectrum of Sine-Gordon theory}},  {\em Nucl. Phys.} {\bf B540}
  (1999) 543--586 [\href{http://arXiv.org/abs/hep-th/9805117}{{\tt
  hep-th/9805117}}].

\bibitem{Zamolodchikov:1994uw}
A.~B. Zamolodchikov, {\it {Painleve III and 2-d polymers}},  {\em Nucl. Phys.}
  {\bf B432} (1994) 427--456 [\href{http://arXiv.org/abs/hep-th/9409108}{{\tt
  hep-th/9409108}}].

\bibitem{Bazhanov:1996aq}
V.~V. Bazhanov, S.~L. Lukyanov and A.~B. Zamolodchikov, {\it {Integrable
  quantum field theories in finite volume: Excited state energies}},  {\em
  Nucl. Phys.} {\bf B489} (1997) 487--531
  [\href{http://arXiv.org/abs/hep-th/9607099}{{\tt hep-th/9607099}}].

\bibitem{DT}
P.~Dorey and R.~Tateo, {\it {Excited states by analytic continuation of TBA
  equations}},  {\em Nucl. Phys. B} {\bf 482} (1996) 639--659
  [\href{http://arXiv.org/abs/hep-th/9607167}{{\tt hep-th/9607167}}].

\bibitem{Destri:1994bv}
C.~Destri and H.~J. De~Vega, {\it {Unified approach to thermodynamic Bethe
  Ansatz and finite size corrections for lattice models and field theories}},
  {\em Nucl. Phys. B} {\bf 438} (1995) 413--454
  [\href{http://arXiv.org/abs/hep-th/9407117}{{\tt hep-th/9407117}}].

\bibitem{Fioravanti:2004cz}
D.~Fioravanti, {\it {Geometrical loci and CFTs via the Virasoro symmetry of the
  mKdV-SG hierarchy: An Excursus}},  {\em Phys. Lett. B} {\bf 609} (2005)
  173--179 [\href{http://arXiv.org/abs/hep-th/0408079}{{\tt hep-th/0408079}}].

\bibitem{Faddeev:1987ph}
L.~D. Faddeev and L.~A. Takhtajan, {\em {Hamitonian methods in the theory of
  solitons}}.
\newblock 1987.

\bibitem{Babelon:2003qtg}
O.~Babelon, D.~Bernard and M.~Talon, {\em {Introduction to Classical Integrable
  Systems}}.
\newblock Cambridge Monographs on Mathematical Physics. Cambridge University
  Press, 2003.

\bibitem{Alday:2009dv}
L.~F. Alday, D.~Gaiotto and J.~Maldacena, {\it {Thermodynamic Bubble Ansatz}},
  {\em JHEP} {\bf 09} (2011) 032
  [\href{http://arXiv.org/abs/arXiv:0911.4708}{{\tt arXiv:0911.4708}}].

\bibitem{Alday:2010vh}
L.~F. Alday, J.~Maldacena, A.~Sever and P.~Vieira, {\it {Y-system for
  Scattering Amplitudes}},  {\em J. Phys.} {\bf A43} (2010) 485401
  [\href{http://arXiv.org/abs/arXiv:1002.2459}{{\tt arXiv:1002.2459}}].

\bibitem{Ito:2015nla}
K.~Ito and C.~Locke, {\it {ODE/IM correspondence and Bethe ansatz for affine
  Toda field equations}},  {\em Nucl. Phys. B} {\bf 896} (2015) 763--778
  [\href{http://arXiv.org/abs/arXiv:1502.00906}{{\tt arXiv:1502.00906}}].

\bibitem{Ito:2016qzt}
K.~Ito and H.~Shu, {\it {ODE/IM correspondence for modified $B_2^{(1)}$ affine
  Toda field equation}},  {\em Nucl. Phys. B} {\bf 916} (2017) 414--429
  [\href{http://arXiv.org/abs/arXiv:1605.04668}{{\tt arXiv:1605.04668}}].

\bibitem{Guica:2017lia}
M.~Guica, {\it {An integrable Lorentz-breaking deformation of two-dimensional
  CFTs}},  {\em SciPost Phys.} {\bf 5} (2018), no.~5 048
  [\href{http://arXiv.org/abs/1710.08415}{{\tt 1710.08415}}].

\bibitem{Chakraborty:2018vja}
S.~Chakraborty, A.~Giveon and D.~Kutasov, {\it {$ J\overline{T} $ deformed
  CFT$_{2}$ and string theory}},  {\em JHEP} {\bf 10} (2018) 057
  [\href{http://arXiv.org/abs/1806.09667}{{\tt 1806.09667}}].

\bibitem{Ferko:2022cix}
C.~Ferko, A.~Sfondrini, L.~Smith and G.~Tartaglino-Mazzucchelli, {\it {Root-$T
  \bar T$ Deformations in Two-Dimensional Quantum Field Theories}},  {\em Phys.
  Rev. Lett.} {\bf 129} (2022), no.~20 201604
  [\href{http://arXiv.org/abs/arXiv:2206.10515}{{\tt arXiv:2206.10515}}].

\bibitem{Babaei-Aghbolagh:2022leo}
H.~Babaei-Aghbolagh, K.~Babaei~Velni, D.~Mahdavian~Yekta and H.~Mohammadzadeh,
  {\it {Marginal $T \bar T$-like deformation and modified Maxwell theories in
  two dimensions}},  {\em Phys. Rev. D} {\bf 106} (2022), no.~8 086022
  [\href{http://arXiv.org/abs/arXiv:2206.12677}{{\tt arXiv:2206.12677}}].

\bibitem{Hou:2022csf}
J.~Hou, {\it {$T\bar{T}$ flow as characteristic flows}},
  \href{http://arXiv.org/abs/arXiv:2208.05391}{{\tt arXiv:2208.05391}}.

\bibitem{Borsato:2022tmu}
R.~Borsato, C.~Ferko and A.~Sfondrini, {\it {On the Classical Integrability of
  Root-$T \overline{T}$ Flows}},
  \href{http://arXiv.org/abs/arXiv:2209.14274}{{\tt arXiv:2209.14274}}.

\bibitem{Tempo:2022ndz}
D.~Tempo and R.~Troncoso, {\it {Nonlinear automorphism of the conformal algebra
  in 2D and continuous $\sqrt{T\bar{T}}$ deformations}},
  \href{http://arXiv.org/abs/arXiv:2210.00059}{{\tt arXiv:2210.00059}}.

\bibitem{Garcia:2022wad}
J.~A. Garc\'\i{}a and R.~A. S\'anchez-Isidro, {\it {$\sqrt{T\bar{T}}$-deformed
  oscillator inspired by ModMax}},
  \href{http://arXiv.org/abs/arXiv:2209.06296}{{\tt arXiv:2209.06296}}.

\bibitem{Dorey:2019ngq}
P.~Dorey, C.~Dunning, S.~Negro and R.~Tateo, {\it {Geometric aspects of the
  ODE/IM correspondence}},  {\em J. Phys. A} {\bf 53} (2020), no.~22 223001
  [\href{http://arXiv.org/abs/arXiv:1911.13290}{{\tt arXiv:1911.13290}}].

\bibitem{Gregori:2022xks}
D.~Gregori and D.~Fioravanti, {\it {Quasinormal modes of black holes from
  supersymmetric gauge theory and integrability}},  {\em PoS} {\bf ICHEP2022}
  (11, 2022) 422.

\bibitem{Dorey:2004fk}
P.~Dorey, A.~Millican-Slater and R.~Tateo, {\it {Beyond the WKB approximation
  in PT-symmetric quantum mechanics}},  {\em J. Phys. A} {\bf 38} (2005)
  1305--1332 [\href{http://arXiv.org/abs/hep-th/0410013}{{\tt
  hep-th/0410013}}].

\bibitem{Conti:2020zft}
R.~Conti and D.~Masoero, {\it {Counting monster potentials}},  {\em JHEP} {\bf
  02} (2021) 059 [\href{http://arXiv.org/abs/arXiv:2009.14638}{{\tt
  arXiv:2009.14638}}].

\end{thebibliography}\endgroup
\end{document}